\documentclass[pre,showpacs,showkeys,preprintnumbers,amsmath,amssymb,superscriptaddress,twocolumn]{revtex4-2}
\usepackage{graphicx}
\usepackage{amssymb,amsfonts,amsmath}
\usepackage{epstopdf}
\usepackage{color}
\usepackage{bm}
\usepackage[colorlinks=true,linkcolor=blue,citecolor=red]{hyperref}%
\usepackage{comment}
\usepackage{multirow}

\def\erfc{\mathrm{erfc}}
\def\erfcx{\mathrm{erfcx}}

\def\n{\bm{n}}

\def\x{\bm{x}}

\def\pa{{\partial \Omega}}
\def\ve{\varepsilon}

\def\kappae{\kappa_{\rm e}}
\def\He{H_{\rm e}}
\def\Heba{H_{\rm e}}
\def\taue{\tau_{\rm e}}

\def\C{\mathbb{C}}

\def\K{\mathcal{K}}
\def\Ke{\K_{\rm e}}
\def\L{{\mathcal L}}
\def\M{{\mathcal M}}

\def\N{\mathcal{N}}
\def\P{{\mathbb P}}
\def\R{{\mathbb R}}

\def\T{\mathcal{T}}

\def\X{\bm{X}}

\begin{document}

\title{Diffusion-controlled reactions with time-dependent surface reactivity}

\author{Denis~S.~Grebenkov}
 \email{denis.grebenkov@polytechnique.edu}

\affiliation{Laboratoire de Physique de la Mati\`{e}re Condens\'{e}e, \\ 
CNRS -- Ecole Polytechnique, Institut Polytechnique de Paris, 91120 Palaiseau, France}

\author{Yilin Ye}
 \email{yyl.libri@gmail.com}

\affiliation{Laboratoire de Physique de la Mati\`{e}re Condens\'{e}e, \\ 
CNRS -- Ecole Polytechnique, Institut Polytechnique de Paris, 91120 Palaiseau, France}

\date{\today}

\begin{abstract}
We investigate diffusion-controlled reactions on a reactive surface
with time-dependent reactivity.  Starting from a conventional
formulation via diffusion equation with Robin boundary condition for
the macroscopic concentration of particles, we aim at quantifying how
temporal variations of reactivity affect the survival probability and
the associated first-reaction time (FRT) of a single particle on the
reactive surface.  Using an adiabatic description, we first identify
five long-time regimes of the first-reaction time distribution,
ranging from exponential and stretched-exponential to power-law decay.
In particular, sufficiently rapid decay of the reactivity leads to a
nonzero probability of never reacting, despite diffusion in a bounded
domain.  Second, we provide a probabilistic reformulation of this
problem and compare it with a different scenario of
encounter-dependent reactivity.  We use a heuristic mapping to compare
the long-time kinetics of time-dependent and encounter-dependent
reactivity.  Our theoretical predictions are validated via extensive
Monte Carlo simulations.
\end{abstract}



\maketitle

\section{Introduction}
\label{sec:intro}

Many biochemical reactions occurring in living cells or industrial
chemical reactors rely on diffusive transport for mixing reactants and
ensuring their encounters \cite{House,Lindenberg}.  Typical examples
range from the diffusive search by proteins for specific sites on DNA
\cite{Phillips,Berg87,Mirny09,Kolomeisky11,Sheinman12,Bressloff13,Shin19}
to heterogeneous catalysis
\cite{Thomas,Coppens99,Barzykin01,Grebenkov19c,Punia21,Grebenkov26c}.
The time evolution of a macroscopic concentration $C$ of diffusing
reactants is generally described by a diffusion (or Fokker-Planck)
equation with suitable boundary conditions accounting for surface
reactions \cite{North66,Wilemski73,Calef83,Berg85,Rice}.  Starting
from the seminal work by Collins and Kimball \cite{Collins49}, a
reactive surface $\Gamma$ with partial reactivity is described by the
Robin boundary condition (also known as radiation condition or
partially absorbing condition),
\begin{equation}  \label{eq:Robin0}
-D \partial_n C = \kappa_0 C, 
\end{equation}
which postulates that the diffusive flux of reactants at each point of
$\Gamma$ (expressed by the normal derivative $\partial_n C$) is
proportional to their concentration on $\Gamma$.  Here the diffusion
coefficient $D$ characterizes the diffusive motion of reactants in the
bulk, whereas the constant reactivity $\kappa_0$ (in units of length
per time) can represent various surface reaction mechanisms on
$\Gamma$ such as overpassing an activation energy or entropic barrier
for the successful reaction event \cite{Zhou91,Reguera06,Chapman16},
microscopic heterogeneity of the reactive surface
\cite{Berg77,Berezhkovskii04,Muratov08,Bernoff18}, dynamic changes of
reactant activity (e.g., via stochastic switching)
\cite{Bashford86,Zhou96,Green96,Benichou00,Reingruber09,Lawley15,Bressloff17},
etc. (see an overview in \cite{Grebenkov23n}).  The relative roles of
diffusive dynamics, partial reactivity, structure of the confinement,
shape and size of the reactive surface have been thoroughly
investigated
\cite{Redner,ben-Avraham,Schuss,Metzler,Dagdug,Sano79,Sapoval94,Grebenkov05,Benichou10,Benichou14,Galanti16,Serov16,Piazza22}.

The assumption of constant reactivity $\kappa_0$, common in earlier
theoretical studies, permits Laplace transformation of the diffusion
equation and facilitates spectral and asymptotic analyses
\cite{Gardiner,Risken,Carslaw,Crank}.  In many applications, however,
the reactive surface does evolve with time.  For instance, if the
reaction event consists of passing through an ion channel, its
effective ``reactivity'' varies as the channel switches between open
and closed states.  In neuronal communication, such switching can be
triggered by changes in the membrane potential induced by an incoming
electrical signal.  Similarly, the reactivity of enzymes is
conditioned by their environment, allowing for their progressive
activation or passivation.  In heterogeneous catalysis, the activity
of catalytic germs may deteriorate over time
\cite{Bartholomew01,Filoche08,Abad15}; since their replacement can
impose substantial costs on the chemical industry, an accurate
description of the effect of variable reactivity poses an important
challenge.

When modeling such systems, one usually assumes that reactivity
modulations are controlled by the intrinsic dynamics of the reactive
surface $\Gamma$ (like the switching dynamics of a channel) that
prescribes a time-dependent reactivity $\kappa(t)$.  At each encounter
with $\Gamma$, the diffusing particle experiences the reactivity,
evaluated at time $t$ of that encounter, independently of how many
encounters occurred up to $t$.  The concentration is thus governed by
the diffusion equation, subject to the Robin boundary condition
(\ref{eq:Robin0}) with $\kappa(t)$ in place of $\kappa_0$.
A different yet related class of diffusion-mediated reactions with
nonexponential reaction kinetics was investigated in \cite{Re00}
within a continuous-time random-walk framework on a lattice.  Here,
the time-dependent absorption rate depends on the elapsed time since
the particle arrived at a reactive site, rather than on the absolute
time $t$ considered in our paper.
Another class of models arises in reversible binding, where the
elimination of bound states can lead to an effective reactivity that
depends on the Laplace variable
\cite{Agmon84,Agmon90,Kim99,Gopich00,Gopich16}.  Such a
frequency-dependent reactivity corresponds to a memory kernel in the
time domain and should be distinguished from the prescribed
instantaneous reactivity $\kappa(t)$ considered here.  In particular,
the former generally leads to a nonlocal boundary condition in time,
whereas the latter retains the local Robin form.
Despite the apparent simplicity of this local form, obtaining an
explicit solution for an arbitrary time-dependent reactivity
$\kappa(t)$ remains challenging, even in simple geometries such as an
interval.  Understanding how the time-dependent reactivity affects the
overall reaction kinetics and the distribution of the first-reaction
time (FRT) on $\Gamma$ is a main motivation of this work.

In some systems, the reactivity is not imposed deterministically but
can change at each encounter of the diffusing particle with the
reactive surface $\Gamma$, thus being controlled by their diffusive
interactions \cite{Grebenkov20}.  The history of such random
encounters is characterized by the boundary local time $\ell_t$ (see
Sec. \ref{sec:encounter}), whereas the reactive properties of $\Gamma$
are prescribed via an encounter-dependent reactivity $\kappae(\ell)$.
In sharp contrast to the time-dependent scenario, the reactivity at
time $t$, $\kappae(\ell_t)$, is not deterministic but randomly
modulated by the diffusive dynamics via the boundary local time
$\ell_t$.  How different are these two scenarios?

In this paper, we investigate both time-dependent and
encounter-dependent scenarios and establish similarities and
differences between their reaction kinetics.  Section
\ref{sec:encounter} recalls the encounter-based approach and analyzes
its long-time asymptotic regimes.  Section \ref{sec:time} develops a
probabilistic formulation of time-dependent reactivity, derives
asymptotic results using an adiabatic approximation, and proposes a
heuristic mapping between two scenarios.  Section
\ref{sec:numerics} presents numerical results, while
Sec. \ref{sec:conclusion} summarizes our findings.  Technical
derivations and numerical methods are provided in Appendices
\ref{sec:Aencounter} -- \ref{sec:Anumerics}.

\section{Encounter-dependent reactivity}
\label{sec:encounter}

In this Section, we first recall the encounter-based approach to
describe the impact of encounter-dependent reactivity onto
diffusion-controlled reactions.  In particular, we refine the
long-time asymptotic analysis of the survival probability from
\cite{Grebenkov20} and provide an equivalent stopping condition.  This
section will serve as a basis for extending the developed concepts to
the time-dependent scenario in Sec. \ref{sec:time}.

\subsection{Encounter-based approach}

\begin{figure}
\centering
\includegraphics[width=0.55\columnwidth]{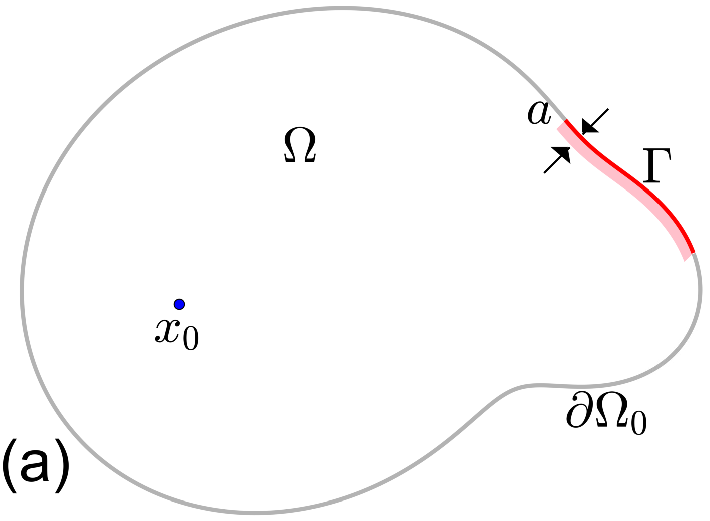} 
\includegraphics[width=0.43\columnwidth]{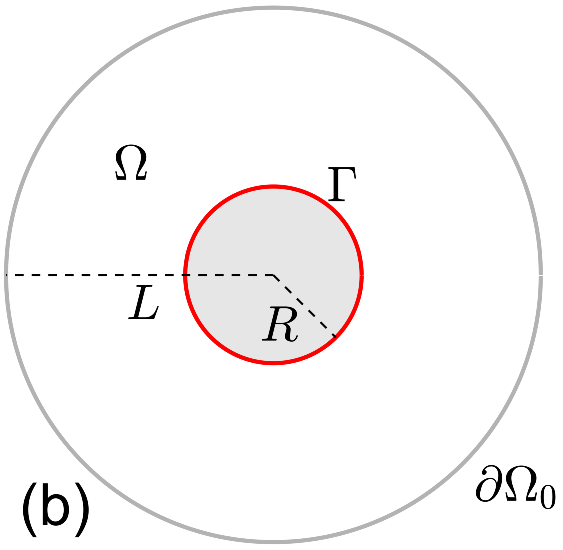} 
\caption{
{\bf (a)} Schematic presentation of a confining domain $\Omega$ whose
smooth boundary $\pa$ is split into a reactive surface $\Gamma$ (red)
and a reflecting wall $\pa_0$ (gray).  A thin reactive boundary layer
of width $a$ near $\Gamma$ is shown in pink.  Note that both $\Gamma$
and $\pa_0$ can be unions of disjoint subsets; moreover, the boundary
$\pa$ does not need to be connected, in particular, reactive surfaces
can be located in the bulk.
{\bf (b)} A spherical shell, whose two concentric spheres of radii $R$
and $L$ represent the reactive surface $\Gamma$ and the reflecting
wall $\pa_0$, respectively. }
\label{fig:scheme}
\end{figure}

We consider diffusion of reactants in a bounded domain $\Omega
\subset\R^d$, whose smooth boundary $\pa$ is split into two disjoint
parts: a reactive surface $\Gamma$ and an inert impermeable boundary
$\pa_0$ (Fig. \ref{fig:scheme}(a)).  In order to describe the
statistics of first-reaction times on $\Gamma$, we rely on the
encounter-based approach \cite{Grebenkov20}, which starts from the
stochastic Skorokhod equation for reflected Brownian motion
(i.e., a Langevin equation accounting for a reflecting boundary)
\cite{Ito,Freidlin}.  
Random encounters of a single diffusing particle with the reactive
surface $\Gamma$ are represented by the boundary local time on
$\Gamma$: $\ell_t = \lim\limits_{a\to 0} a\, \N_t^{(a)}$.  Here
$\N_t^{(a)}$ is the random number of encounters with a thin layer
$\Gamma_a$ of width $a$, adjacent to $\Gamma$, up to time $t$.
Despite its name, the boundary local time has units of length.
Alternatively, the boundary local time can be defined as the rescaled
residence time in $\Gamma_a$,
\begin{equation}
\ell_t = \lim\limits_{a\to0} \frac{D}{a} \int\limits_0^t dt'\, 1_{\Gamma_a}(\X_{t'}), 
\end{equation}
where $D > 0$ is a constant diffusion coefficient, $\X_{t'}$ denotes
reflected Brownian motion in $\Omega$, and $1_{\Gamma_a}(\x)$ is the
indicator function of $\Gamma_a$: $1_{\Gamma_a}(\x) = 1$ if
$\x\in\Gamma_a$, and $0$ otherwise.

At each encounter with $\Gamma_a$, the particle either reacts
with a small probability $p_{\rm e}$, or resumes its diffusion outside
the boundary layer with probability $1-p_{\rm e}$ (similar concepts
appear in the microscopic description of permeable barriers, see
\cite{Bressloff22,Bressloff23,Kay22,Kay25}).  The probability of
reaction depends on the time $t$ of the encounter, $p_{\rm e} =
a\kappae(\ell_t)/D$, through a given nonnegative function
$\kappae(\ell)$ --- encounter-dependent reactivity --- that prescribes
how the reactivity varies between successive encounters.  While the
layer width $a$ may have a proper physical or chemical interpretation
(e.g., as a characteristic lengthscale of short-ranged interactions
between the particle and the catalytic surface, typically a few
nanometers), the encounter-based approach gets rid of $a$ by taking
the limit $a\to 0$.

If all reaction attempts are independent of each other, the successful
reaction occurs at a random time
\begin{equation} \label{eq:tau_def}
\taue = \inf\{ t > 0 ~:~ \ell_t > \hat{\ell} \}
\end{equation}
when the boundary local time $\ell_t$ (i.e., the rescaled number of
failed attempts) exceeds an independent random reaction threshold
$\hat{\ell}$, whose distribution is directly determined by the
encounter-dependent reactivity \cite{Grebenkov20}:
\begin{equation}  \label{eq:Psi_kappa}
\Psi(\ell) = \P\{ \hat{\ell} > \ell\} = \exp\left(-\Ke(\ell)\right), 
\end{equation}
where
\begin{equation} \label{eq:Kdef}
\Ke(\ell) = \frac{1}{D} \int\limits_0^\ell d\ell' \, \kappae(\ell')
\end{equation}
(throughout, we assume that $\kappae(\ell)$ is locally integrable).

The stopping condition (\ref{eq:tau_def}) separates diffusive dynamics
(characterized by the boundary local time $\ell_t$) from surface
reactions (encoded in the threshold $\hat{\ell}$).  For constant
reactivity, $\kappae(\ell)=\kappa_0$, this threshold is exponentially
distributed, $\Psi_0(\ell) = e^{-\kappa_0 \ell/D}$, and the
conventional Robin boundary condition (\ref{eq:Robin0}) is recovered
\cite{Grebenkov20}.  In this setting, $\langle \hat{\ell}\rangle
= D/\kappa_0$ is the Robin reaction length.  Conversely, any strictly
positive, absolutely continuous survival function $\Psi(\ell)$ (with
$\Psi(0)=1$) corresponds to an encounter-dependent reactivity, which
is obtained by inverting Eq. (\ref{eq:Psi_kappa}) and using
Eq. (\ref{eq:Kdef}):
\begin{equation}  \label{eq:kappa_Psi}
\kappae(\ell) = - D \frac{\partial_\ell \Psi(\ell)}{\Psi(\ell)}  \,.
\end{equation}
The mathematical description of such more sophisticated surface
reactions is thus not reducible to the conventional framework of the
Robin boundary condition (see an example of non-Markovian binding
kinetics in \cite{Grebenkov23b}).

Despite the complexity of this general setting, the probabilistic
definition (\ref{eq:tau_def}) of the FRT $\taue$ leads to an
analytically tractable theory of diffusion-controlled reactions with
encounter-dependent reactivity \cite{Grebenkov20}.
In fact, let us introduce the first-crossing time $\T_\ell =
\inf\{ t > 0 ~:~ \ell_t > \ell\}$ of a fixed threshold $\ell$ by the
boundary local time $\ell_t$.  The probability density function (PDF)
of the random variable $\T_\ell$, $U(\ell,t)$, depends exclusively on
the diffusive dynamics in $\Omega$ and on the encounters with
$\Gamma$.  According to the stopping condition (\ref{eq:tau_def}), one
has $\taue = \T_{\hat{\ell}}$, where a fixed threshold $\ell$ is
replaced by the random threshold $\hat{\ell}$.  As a consequence, the
PDF of the FRT $\taue$ can be written as the average of $U(\ell,t)$
over all random realizations of the threshold $\hat{\ell}$:
\begin{equation}  \label{eq:Ht_Ut}
\He(t) = \langle U(\hat{\ell},t)\rangle = \int\limits_0^\infty d\ell \,\psi(\ell) \, U(\ell,t) ,
\end{equation}
where $\psi(\ell) = - \partial_\ell \Psi(\ell)$ is the PDF of the
threshold $\hat{\ell}$ (as discussed below, $\hat{\ell}$ is allowed to
be infinite so that $\psi(\ell)$ may not be normalized to $1$).  The
key relation (\ref{eq:Ht_Ut}) reveals the relative effects of the
geometry [represented via $U(\ell,t)$] and chemistry [represented via
$\psi(\ell)$] onto the FRT statistics.
While the distributions of random quantities $\ell_t$, $\T_\ell$, and
$\taue$ depend on the starting point of the particle
\cite{Grebenkov20c,Grebenkov22a}, we focus throughout the manuscript
on the common situation when the starting point is uniformly
distributed in $\Omega$; consequently, the angular brackets denote
averaging over both the reflected Brownian trajectories and the
uniform distribution of the starting point.

\subsection{Equivalent stopping condition}

In view of comparison to the time-dependent scenario below, it is
instructive to provide an equivalent form of the stopping condition
(\ref{eq:tau_def}).  For this purpose, we introduce a reaction
functional:
\begin{equation}  \label{eq:barLt_def}
\bar{L}_t = \Ke(\ell_t) = \frac{1}{D} \int\limits_0^{\ell_t} d\ell' \, \kappae(\ell').
\end{equation}
Since the encounter-dependent reactivity $\kappae(\ell)$ is
nonnegative, the process $\bar{L}_t$ is nondecreasing.  Introducing an
independent random variable $\hat L$ with the standard exponential
distribution, we can equivalently express the first-reaction time as
\begin{equation}  \label{eq:tau_barL}
\taue = \inf\{ t > 0 ~:~ \bar{L}_t > \hat{L} \}.
\end{equation}
Indeed, conditionally on the trajectory of reflected Brownian motion,
the probability of surviving up to time $t$ is $\exp[-\Ke(\ell_t)]$,
which coincides with $\Psi(\ell_t)$ in Eq. (\ref{eq:Psi_kappa}); in
particular, the survival probability of the particle is
\begin{equation}  \label{eq:Se_barLt}
S_{\rm e}(t) = \P\{ \taue > t\} = \P\{ \bar{L}_t < \hat{L} \} =  \langle e^{-\bar{L}_t}\rangle .
\end{equation}
This representation remains valid even when $\Ke(\ell)$ is not
strictly increasing, or if $\Ke(\infty)<\infty$, with the convention
$\inf \varnothing =\infty$.  In the special case of a constant
reactivity, $\kappae(\ell) = \kappa_0$, the expectation $\langle
e^{-\bar{L}_t}\rangle$ with $\bar{L}_t = \kappa_0 \ell_t/D$ is the
Feynman-Kac representation of the Robin problem.  The encounter-based
approach can thus be seen as its generalization to a reactivity that
depends on accumulated encounters.

While the stopping condition (\ref{eq:tau_def}) separates diffusion
from surface reactions, the equivalent formulation (\ref{eq:tau_barL})
is better suited for comparison with time-dependent reactivity in
Sec. \ref{sec:time}.

\subsection{Asymptotic behavior of the FRT distribution}
\label{sec:regimes}

The fundamental relation (\ref{eq:Ht_Ut}) has numerous practical
consequences.  Notably, it reveals how the behavior of the
encounter-dependent reactivity $\kappae(\ell)$ at large $\ell$
determines the asymptotic decay of $\He(t)$ at long times.  In
particular, if the encounter-dependent reactivity behaves as
$\kappa_e(\ell)\propto\ell^{-1-\alpha}$, we distinguish five
asymptotic regimes (see Appendix \ref{sec:Along}):
\begin{equation}  \label{eq:He_power0}
\He(t) \simeq \begin{cases} 
\exp\bigl(-Dt\lambda_0^{\infty} + o(t)\bigr) & (\alpha < -1), \cr
\exp\bigl(-Dt\lambda_0^{\kappa_\infty} + o(t)\bigr) & (\alpha = -1), \cr
\exp\bigl(-c\, t^{|\alpha|} + o(t^{|\alpha|})\bigr) & (-1< \alpha < 0), \cr
A_0 t^{-\nu-1}  & (\alpha = 0), \cr
A_\alpha t^{-\alpha-1}  & (\alpha > 0)  \end{cases}
\end{equation}
(the first line in Eq. (\ref{eq:He_power0}) is also valid if
$\kappae(\ell)$ grows faster than a power law).  Here $c$, $\nu$ and
$A_\alpha$ are constants depending on finer details of
$\kappae(\ell)$, $\kappa_\infty$ is the limiting value of
$\kappae(\ell)$ in the case $\alpha = -1$, whereas
$\lambda_0^{\infty}$ [resp.  $\lambda_0^{\kappa_\infty}$] is the
principal eigenvalue of the Dirichlet [resp., Robin] Laplacian, see
Eq. (\ref{eq:Robin_eigen}) below for a formal definition.
In the exponential and stretched-exponential regimes, these relations
specify the leading term of $\ln \He(t)$, while subleading corrections
in the exponent may remain significant.  In the power-law regimes, the
expressions specify the decay exponents, whereas the corresponding
multiplicative constants, possible logarithmic or other slowly varying
corrections depend on the reaction kinetics and the geometry.

Let us comment on various asymptotic regimes.

(i) The value $\alpha = -1$ separates three distinct behaviors of the
encounter-dependent reactivity: asymptotically growing ($\alpha <
-1$), constant ($\alpha = -1$), and decreasing ($\alpha > -1$).  The
first two regimes lead to an exponential decay of $\He(t)$, as in the
case of a constant reactivity; in particular, all the moments of the
FRT $\taue$ are finite.  In contrast, the decreasing reactivity
results in an atypical, non-exponential decay of $\He(t)$.

(ii) In the critical regime $\kappae(\ell) \simeq \nu D/\ell$ at large
$\ell$ (for some dimensionless $\nu > 0$), the PDF $\He(t)$ decays as
$t^{-1-\nu}$ (up to a slowly varying factor), whose exponent $\nu$ is
determined by the amplitude of the reactivity.  In particular, the
mean FRT is finite for $\nu > 1$ and infinite for $\nu < 1$ (at $\nu =
1$, its finiteness depends on subleading terms in $\kappa_e(\ell)$).

(iii) If $\kappae(\ell)$ decays as $\ell^{-1-\alpha}$ with an exponent
$\alpha > 0$, its integral $\Ke(\ell)$ in Eq. (\ref{eq:Kdef}) remains
finite even in the limit $\ell\to \infty$, so that the FRT can be
infinite with a strictly positive probability:
\begin{equation}  \label{eq:Pe_infty}
\P\{ \taue = \infty\} = \P\{ \hat{\ell} = \infty\} 
= \Psi(\infty) = e^{- \Ke(\infty)} > 0.
\end{equation}
This is the probability of no surface reaction in a bounded domain:
even though the diffusive exploration is compact, the reactivity
decays fast enough so that the particle may fail to react even after
an infinite number of returns to the reactive surface.

(iv) The emergence of the principal Dirichlet eigenvalue
$\lambda_0^\infty$ for $\alpha<-1$ deserves some explanation.  The
Dirichlet boundary condition describes perfect absorption upon the
first encounter with $\Gamma$, in which case the long-time survival is
governed by trajectories that have avoided $\Gamma$.  For
time-dependent reactivity, this interpretation is natural: as
$\kappa(t)\to\infty$, late encounters become increasingly likely to
cause a reaction, leading to the Dirichlet decay rate (see
Sec. \ref{sec:adiabatic}).  In contrast, for encounter-dependent
reactivity, a particle that has avoided $\Gamma$ has accumulated no
boundary local time, and its reactivity has not increased.  Thus, the
appearance of the Dirichlet eigenvalue in Eq. (\ref{eq:He_power0}) is
less intuitive and reflects a nontrivial property of the
encounter-dependent dynamics.

(v) As discussed in Appendix \ref{sec:Along}, subleading corrections
in the exponential or stretched-exponential functions can be
significant at moderately long times, even if they do not alter the
leading logarithmic asymptotics.

For asymptotically decreasing reactivity ($\alpha > - 1$), the
analysis in Appendix \ref{sec:Along} further suggests that the
long-time decay of $\He(t)$ is closely related to that of the
threshold density $\psi(\ell)$, evaluated at $\ell = vt$, where
\begin{equation}  \label{eq:v_def}
v = D\frac{|\Gamma|}{|\Omega|} \,, 
\end{equation}
with $|\Gamma|$ and $|\Omega|$ denoting respectively the surface area
of $\Gamma$ and the volume of $\Omega$.
The opposite short-time behavior is determined by $\psi(\ell)$ near
$\ell=0$ (Appendix \ref{sec:Ashort}).  In particular, if $\psi(0)>0$
is finite, then
\begin{equation}
\He(t) \simeq v \, \psi(0)  \qquad (t\to 0).
\end{equation}
Motivated by these two limiting behaviors, we propose the following
simple interpolation for intermediate times:
\begin{equation}  \label{eq:Ht_approx}
\He(t) \approx v \, \psi(v t)  \qquad (t \geq 0).
\end{equation}
Using Eq. (\ref{eq:Psi_kappa}), we can also write this approximation
in terms of the encounter-dependent reactivity:
\begin{equation}  \label{eq:Ht_approx2}
\He(t) \approx \frac{|\Gamma|}{|\Omega|} \kappae(vt) e^{-\Ke(vt)} .
\end{equation}
For constant and sufficiently small reactivity,
Eq. (\ref{eq:Ht_approx2}) recovers the classical reaction-limited
exponential distribution, with mean FRT $|\Omega|/(\kappa_0|\Gamma|)$
\cite{Sano79,Szabo80}.  This approximation is generally not valid at
high reactivity, illustrating the failure of Eq. (\ref{eq:Ht_approx2})
for $\alpha \leq -1$.

\section{Time-dependent reactivity}
\label{sec:time}

This section provides our main results for the time-dependent
scenario.  After recalling the macroscopic description of this
problem, we employ an adiabatic description to obtain the long-time
behavior of the PDF $H(t)$.  Using a probabilistic reformulation, we
explore similarities and differences between the time-dependent and
encounter-dependent scenarios.

\subsection{Conventional formulation}

A conventional formulation of time-dependent reactivity starts from
the macroscopic concentration $C(\x,t)$ of reactants satisfying the
diffusion equation,
\begin{equation}   \label{eq:diffusion}
\partial_t C = D \Delta C  \quad \textrm{in}~\Omega ,
\end{equation}
where $\Delta$ is the Laplace operator.  This equation describes time
evolution of the concentration from an initial concentration at time
$0$ that for simplicity we assume to be homogeneous: $C(\x,0) = C_0$.
As the impermeable boundary just confines the particles within the
domain $\Omega$, the concentration obeys no-flux (Neumann) boundary
condition on $\pa_0$:
\begin{equation}  \label{eq:Neumann}
\partial_n C = 0  \quad \textrm{on}~\pa_0,   
\end{equation}
where $\partial_n$ is the normal derivative oriented outward from
$\Omega$.

On the reactive surface $\Gamma$, the diffusive flux is balanced by
the reaction flux, yielding the time-dependent Robin boundary
condition
\begin{equation}  \label{eq:Robin}
-D \partial_n C = \kappa(t) \, C  \quad \textrm{on}~ \Gamma .
\end{equation}
This is an immediate extension of the classical Collins-Kimball's mass
conservation law \cite{Collins49}, in which the prescribed reactivity
$\kappa(t)$ characterizes the intrinsic dynamics of the reactive
surface, independently of the diffusive motion of reactants in
$\Omega$.
The reactivity may range from zero, corresponding to a perfectly
reflecting surface, to the formal limit $\kappa(t)\to\infty$,
corresponding to perfect absorption.  We assume that $\kappa(t)$ is
nonnegative and sufficiently regular for the diffusion problem
(\ref{eq:diffusion} -- \ref{eq:Robin}) to be well posed
\cite{Lions,Dautray,Friedman}.  The probabilistic formulation in
Sec. \ref{sec:time_proba} also accommodates nonnegative, locally
bounded reactivities, including piecewise-constant switching
protocols; in turn, the adiabatic analysis of Sec. \ref{sec:adiabatic}
requires additional smoothness.  Once the solution is found, the total
reaction rate of this system is characterized by the diffusive flux of
reactants on $\Gamma$ at time $t$:
\begin{equation} \label{eq:J_def}
J(t) = \int\limits_{\Gamma} d\x \, (-D\partial_n C(\x,t)) . 
\end{equation}

Diffusion-controlled reactions with time-dependent absorption rates
have been investigated in earlier works, including continuous-time
random-walk models and stochastic descriptions of dynamically reactive
boundaries \cite{Bashford86,Zhou96,Green96,Re00}.  Closely related
mathematical problems arise in transient heat conduction with
time-dependent convective boundary conditions, for which numerical
methods and analytical solutions in simple geometries have been
developed \cite{Ozisik74,Becker83,Chen10,Chiba18}.  In
particular, a standard time-dependent Green's formula allows one to
transform the above PDE to an equivalent integral equation
\begin{align} \label{eq:C_integral}
C(\x,t) & = \int\limits_{\Omega} d\x' \, C_0\, G_0(\x',t|\x,0) \\  \nonumber
& - \int\limits_0^t dt' \kappa(t') \int\limits_{\Gamma} d\x' \, C(\x',t')\, G_0(\x',t-t'|\x,0),
\end{align}
where $G_0(\x',t|\x,0)$ is the diffusion propagator (or heat kernel)
with reflecting boundary condition on the whole boundary, including
$\Gamma$.  Restricting Eq. (\ref{eq:C_integral}) to $\x\in\Gamma$
gives a Volterra integral equation for the boundary concentration.
For a planar boundary with tangentially uniform data, or for a
concentric spherical shell with radially symmetric data, the boundary
concentration is spatially uniform and the equation reduces to a
scalar time convolution.  In this simplified setting, the correction
term has the same Green's function structure as the history term in
the Ward-Tordai description of adsorption \cite{Ward46,Delahay57},
with the prescribed Robin sink $\kappa(t) C|_\Gamma$ providing the
surface flux here.
However, for arbitrary geometry, the time-dependent Robin condition
couples the spatial eigenmodes, preventing a direct application of the
standard spectral solution for constant reactivity.

Moreover, an extension of the above macroscopic description to a
single-particle stochastic description remains elusive.  In the case
of a constant reactivity, a macroscopic concentration, after rescaling
by the uniform initial concentration $C_0$,
\begin{equation}  \label{eq:St_Ct}
S(t|\x) = \frac{C(\x,t)}{C_0} \,,
\end{equation}
can be interpreted as the probability that a single particle started
from a point $\x$ at time $0$ has not reacted on $\Gamma$ by time $t$
\cite{Redner}.  This survival probability defines the probability law
of a random variable $\tau_{\x}$, $S(t|\x) = \P\{ \tau_{\x} > t\}$,
that can be interpreted as the first-reaction time of that particle on
$\Gamma$.
However, an extension of this construction to time-dependent
reactivity $\kappa(t)$ is not straightforward.  In fact, the time
dependence in $\kappa(t)$ breaks the symmetry between the starting and
arrival points.  While the concentration satisfies the forward
diffusion equation, the survival probability obeys the backward one
\cite{VanKampen} so that the pointwise relation (\ref{eq:St_Ct}) does
not hold anymore.  In this more general case, one can start from the
backward diffusion equation or, equivalently, introduce the survival
probability by integrating the propagator (or heat kernel)
$G(\x,t|\x_0,0)$ over the arrival point $\x$:
\begin{equation}  \label{eq:Sx0_def}
S(t|\x_0) = \int\limits_{\Omega} d\x \, G(\x,t|\x_0,0).
\end{equation}
This is a continuous monotonically decreasing function of time $t$
that defines a nonnegative, possibly infinite random variable
\cite{Feller}, which can be interpreted as the first-reaction time.

In the case of the uniform starting point, the equivalence between the
concentration and the survival probability is recovered.  In fact, one
has
\begin{align} \nonumber
C(t) & = \frac{1}{|\Omega|} \int\limits_{\Omega} d\x \, C(\x,t) = 
\int\limits_{\Omega} \frac{d\x}{|\Omega|}  \int\limits_{\Omega} d\x_0\, G(\x,t|\x_0,0) \, C_0 \\  \label{eq:Ct_St}
& = \frac{C_0}{|\Omega|} \int\limits_{\Omega} d\x_0 \, S(t|\x_0) = C_0 \, S(t).
\end{align}
Using Eq. (\ref{eq:J_def}) and the divergence theorem, the time
derivative of Eq. (\ref{eq:Ct_St}) yields the PDF of the FRT $\tau$:
\begin{equation}  \label{eq:Ht_Jt}
H(t) = -\partial_t S(t) = \frac{J(t)}{C_0 |\Omega|} \,. 
\end{equation}
At short times, Eq. (\ref{eq:J_def}) implies
\begin{equation}
H(t) \approx \frac{|\Gamma|}{|\Omega|} \kappa(t) \qquad (t\to 0).  
\end{equation}
In turn, the behavior of $H(t)$ is much more sophisticated at moderate
and long times, as discussed below.

Although Eq. (\ref{eq:Sx0_def}) or Eq. (\ref{eq:Ct_St}) formally
defines the FRT distribution, it does not provide an explicit stopping
condition in terms of single-particle stochastic dynamics.  Such a
condition is established in Sec. \ref{sec:time_proba} by extending the
encounter-based formulation to time-dependent reactivity.  Before
proceeding to the probabilistic reformulation, we analyze the
long-time behavior of the survival probability $S(t)$ in
Secs. \ref{sec:adiabatic} and \ref{sec:Ht_regimes}.

\subsection{Adiabatic approximation}
\label{sec:adiabatic}

For a constant reactivity $\kappa$, the solution of the time-dependent
problem (\ref{eq:diffusion}, \ref{eq:Neumann}, \ref{eq:Robin}) admits
a spectral expansion over the eigenvalues and eigenfunctions of the
Robin(-Neumann) Laplacian:
\begin{subequations}  \label{eq:Robin_eigen}
\begin{align}
-\Delta U_j^{\kappa} & = \lambda_j^{\kappa} U_j^{\kappa} \quad \textrm{in}~ \Omega, \\
D\partial_n U_j^{\kappa} + \kappa U_j^{\kappa} & = 0 \quad \textrm{on}~ \Gamma, \\
\partial_n U_j^{\kappa} & = 0  \quad \textrm{on}~ \pa_0 ,
\end{align}
\end{subequations}
with the reactivity $\kappa$ standing in the superscript.  For a
bounded domain $\Omega$, the spectrum is discrete, the eigenvalues can
be enumerated by an index $j = 0,1,\ldots$ in an increasing sequence,
whereas the eigenfunctions form a complete orthonormal basis of
$L^2(\Omega)$ \cite{Levitin}.  In particular, the long-time asymptotic
behavior is determined by the principal (smallest) eigenvalue
$\lambda_0^{\kappa}$, which is simple and is strictly positive for
$\kappa > 0$: $H(t) \propto e^{-Dt\lambda_0^{\kappa}}$.

When the reactivity is time-dependent, the standard spectral expansion
can be generalized as an expansion of the rescaled concentration
$C(\x,t)/C_0$ over the instantaneous basis (see, e.g.,
\cite{Ozisik74}):
\begin{equation} \label{eq:S_expansion}
\frac{C(\x,t)}{C_0} = \sum\limits_{j=0}^\infty c_j(t) \, U_j^{\kappa(t)}(\x).
\end{equation}
Substituting this expansion into the diffusion equation
(\ref{eq:diffusion}) and projecting onto an instantaneous
eigenfunction $U_n^{\kappa(t)}$ yields an infinite system of ordinary
differential equations for the unknown coefficients $c_j(t)$:
\begin{equation}  \label{eq:cn_eqs}
\dot{c}_n(t) = - D \lambda_n^{\kappa(t)} c_n(t) - \dot{\kappa}(t) 
\sum\limits_{j=0 \atop j\ne n}^\infty \bigl(U_n^{\kappa(t)}, \partial_\kappa U_j^{\kappa(t)}\bigr)_{L^2(\Omega)} c_j(t) ,
\end{equation}
where dot denotes time derivative, e.g., $\dot{c}_n = \partial_t c_n$,
and the diagonal term $j=n$ is excluded from the sum due to the
normalization condition $\|U_n^\kappa\|_{L^2(\Omega)} = 1$.  Note that
the Laplacian acts on the instantaneous eigenfunctions
$U_j^{\kappa(t)}$, while the derivative $\partial_\kappa
U_j^{\kappa(t)}$ arises only from differentiating the basis with
respect to time.  We assume here that $\kappa(t)$ is sufficiently
smooth and that the instantaneous eigenfunctions are chosen
differentiably with respect to $\kappa$ (in the presence of degenerate
eigenvalues, the expansion can instead be formulated using the
corresponding spectral projections).
For instance, if $\kappa(t)$ is constant, one has $\dot{\kappa}(t) =
0$, and Eq. (\ref{eq:cn_eqs}) is reduced to $\dot{c}_n(t) + D
\lambda_n^{\kappa} c_n = 0$, thus recovering the standard spectral
expansion with $c_n(t) = c_n(0) e^{-Dt\lambda_n^{\kappa}}$.  In
contrast, time-dependent reactivity makes an explicit solution of this
system intractable.

At long times, an adiabatic approximation can be developed when the
reactivity varies sufficiently slowly and the instantaneous principal
eigenmode remains well separated from the higher modes.  The relevant
ingredients are the spectral gap and the coupling between
instantaneous eigenfunctions, which is proportional to the rate of
variation of the reactivity.  Neglecting the coupling of the principal
mode to the higher modes in Eq. (\ref{eq:cn_eqs}) yields the
approximation:
\begin{equation}  \label{eq:c0_approx}
c_0(t) \approx c_0(t_*) \exp\biggl(-\int\limits_{t_*}^t dt' \, D\lambda_0^{\kappa(t')} \biggr),
\end{equation}
where $t_*$ is a timescale, above which the adiabatic approximation
holds; its range of validity is discussed in Appendix
\ref{sec:Aadiabatic}.  As a consequence, the long-time behavior of the
volume-averaged survival probability reads
\begin{equation}  \label{eq:St_asympt}
S(t) \approx c_0(t_*) \exp\biggl(-\int\limits_{t_*}^t dt' \, D\lambda_0^{\kappa(t')} \biggr) 
\int\limits_{\Omega} \frac{d\x}{|\Omega|} \, U_0^{\kappa(t)}(\x).
\end{equation}
In the present form, this asymptotic relation looks rather formal, as
it includes several yet unknown elements such as $t_*$, $c_0(t_*)$,
$\lambda_0^{\kappa}$ and $U_0^{\kappa}$.  In the following, we
consider separately three cases of asymptotically decreasing,
constant, and increasing reactivity, for which more explicit relations
will be derived.

\subsubsection{Asymptotically decreasing reactivity}

We first investigate the case when the reactivity $\kappa(t)$
asymptotically vanishes at long times.  In the limit $\kappa\to 0$,
one has
\begin{subequations}
\begin{align}  \label{eq:lambda0_small}
\lambda_0^{\kappa} & \approx \frac{|\Gamma|}{D|\Omega|} \kappa + O(\kappa^2), \qquad
U_0^{\kappa} \approx \frac{1}{\sqrt{|\Omega|}} + O(\kappa), \\
\lambda_j^{\kappa} & \to \lambda_j^{0} > 0 \quad (j=1,2,\ldots). 
\end{align} 
\end{subequations}
Substitution of Eq. (\ref{eq:lambda0_small}) without the subleading
$O(\kappa^2)$ term into Eq. (\ref{eq:St_asympt}) gives the long-time
decay
\begin{equation}  \label{eq:St_asympt_k0}
S(t) \approx A \, e^{-\K(t)}  \qquad (t\to \infty), 
\end{equation}
where $A = c_0(t_*) e^{\K(t_*)}/\sqrt{|\Omega|}$ due to replacement of
$U_0^{\kappa(t)}$ by $U_0^{0} = 1/\sqrt{|\Omega|}$, and
\begin{equation}  \label{eq:Kint}
\K(t) = \frac{|\Gamma|}{|\Omega|} \int\nolimits_0^t dt' \, \kappa(t').
\end{equation}
The time derivative of Eq. (\ref{eq:St_asympt_k0}) yields the
adiabatic prediction
\begin{equation}   \label{eq:Ht_asympt2}
H(t) \approx A\, H_{\rm hom}(t)   \qquad (t\to \infty),
\end{equation}
where
\begin{equation}  \label{eq:Hhom}
H_{\rm hom}(t) = \frac{|\Gamma|}{|\Omega|}\kappa(t) \, e^{- \K(t)} \,.
\end{equation}

When $\K(t)\to\infty$ as $t\to\infty$, Eqs. (\ref{eq:St_asympt_k0},
\ref{eq:Ht_asympt2}) capture the leading logarithmic decay under the
assumptions of the adiabatic approximation, e.g., $\ln S(t)\sim
-\K(t)$.  However, the neglected term $O(\kappa^2)$ in
Eq. (\ref{eq:lambda0_small}) gives a subleading correction
$O\bigl(\int\nolimits_{t_*}^t dt' \, \kappa^2(t')\bigr)$, which may
diverge at large $t$ and thus significantly affect the leading-order
contribution in Eqs. (\ref{eq:St_asympt_k0}, \ref{eq:Ht_asympt2}).
Therefore, the approximate amplitude $A$, which incorporates
modulations of the reactivity over a broad range of times up to $t_*$,
is not an exact asymptotic proportionality constant.
Moreover, if $\K(\infty)<\infty$, Eq. (\ref{eq:St_asympt_k0}) provides
only an estimate of the limiting survival probability $S(\infty)$, as
discussed below.

In the particular situation when $\kappa(t)$ is small at all times,
the principal mode relaxes slowly from the very beginning, so that the
timescale $t_*$ can be set to $0$, whereas $c_0(0) \approx (1,
U_0^{0})_{L^2(\Omega)} = \sqrt{|\Omega|}$ and thus $A = 1$.  In this
case, the asymptotic form (\ref{eq:Hhom}) is an adiabatic
low-reactivity approximation of $H(t)$ at all times.  Note that this
approximate result could be easily derived by integrating
Eq. (\ref{eq:diffusion}) over $\x\in\Omega$,
\begin{align*}
& |\Omega| \partial_t C(t) \approx \partial_t  \int\limits_{\Omega} d\x \, C(\x,t) 
= \int\limits_{\Omega} d\x \,D \Delta C(\x,t) = \\ 
& \int\limits_{\Gamma} d\x \, D \partial_n C(\x,t)
= -\kappa(t) \int\limits_{\Gamma} d\x \, C(\x,t) \approx - \kappa(t) |\Gamma| C(t),
\end{align*}
where we used $C(\x,t) \approx C(t)$ due to rapid diffusion (as
compared to reaction kinetics), so that
\begin{equation} \label{eq:Chom}
C(t) \approx C_{\rm hom}(t) = C_0 e^{- \K(t)},
\end{equation}
from which Eq. (\ref{eq:Hhom}) follows immediately.  In this so-called
reaction-limited regime, $C(t)$, $S(t)$, and $H(t)$ do not depend on
the diffusive transport and thus on $D$, as in a well-stirred
spatially homogeneous setting.  The only information about the
confining domain and the reactive surface that is relevant here, is
the surface-to-volume ratio $|\Gamma|/|\Omega|$.  For a constant
(small) reactivity, $\kappa(t) = \kappa_0$, one retrieves the
well-known exponential law for the FRT, with the mean $\langle \tau
\rangle \approx |\Omega|/(|\Gamma|\kappa_0)$ (see \cite{Grebenkov19b}
and references therein).  Note that the adiabatic approximation
resembles the approximation (\ref{eq:Ht_approx2}) for the
encounter-dependent scenario.

\subsubsection{Asymptotically constant reactivity}

When $\kappa(t)\to \kappa_\infty > 0$ as $t\to \infty$, the principal
eigenvalue $\lambda_0^{\kappa(t)}$ does not vanish but approaches
$\lambda_0^{\kappa_\infty} > 0$.  With the help of the approximation
$\kappa(t) \approx \kappa_\infty$ for $t \geq t_*$ with $t_*$ large
enough, Eq. (\ref{eq:St_asympt}) yields
\begin{equation}  \label{eq:St_Robin}
\ln S(t) \simeq -D\lambda_0^{\kappa_\infty} t + o(t) \qquad (t\to \infty).
\end{equation}
Expectedly, we retrieve the exponential decay of the survival
probability, with the rate controlled by the principal eigenvalue of
the Robin Laplacian, as in the case of constant reactivity.  The
subleading correction $o(t)$ accounts for the transient behavior due
to variations of the reactivity $\kappa(t)$ up to $t_*$.

\subsubsection{Asymptotically growing reactivity}

Finally, we consider the case of time-dependent reactivity that grows
to infinity: $\kappa(t) \to \infty$.  In this case, the spectrum of
the Robin Laplacian converges to that of the Dirichlet Laplacian; in
particular, one has
\begin{equation}  \label{eq:lambda_Dirichlet}
\lambda_j^{\kappa} = \lambda_j^{\infty} + O(1/\kappa) , \qquad
U_j^{\kappa} = U_j^{\infty} + O(1/\kappa) .
\end{equation}
Substituting these relations into Eq. (\ref{eq:St_asympt}) and
neglecting subleading terms, we obtain the asymptotic relation similar
to Eq. (\ref{eq:St_Robin}):
\begin{equation}  \label{eq:St_Dirichlet}
\ln S(t) \simeq -D\lambda_0^{\infty} t + o(t)  \qquad (t\to\infty).
\end{equation}
We emphasize again that this asymptotic relation accounts only for the
leading contribution.  According to Eq. (\ref{eq:lambda_Dirichlet}),
the next-order correction is expected to be
$O\bigl(\int\nolimits_{t_*}^t dt'/\kappa(t')\bigr)$.  If $\kappa(t)$
grows slowly as $t^{\gamma}$ (with $0< \gamma < 1$), this correction
grows as $t^{1-\gamma}$: though it remains subleading as compared to
$t$, its contribution may still be significant.

\subsection{Asymptotic regimes}
\label{sec:Ht_regimes}

In order to illustrate different asymptotic regimes described above,
we choose a power-law model for time-dependent reactivity:
\begin{equation}  \label{eq:kappa_model}
\kappa(t) = \kappa_0 (1+t/t_0)^{-\alpha-1} ,
\end{equation}
which has three parameters: the initial reactivity $\kappa_0$, the
timescale $t_0$, and the exponent $\alpha$ allowing to represent
decreasing, constant, and increasing reactivity.  Note that $\alpha =
-1$ includes the constant reactivity case as a benchmark.  Within this
model, we have
\begin{equation}  \label{eq:K_model}
\K(t) = \begin{cases} \frac{\nu}{\alpha} \bigl(1 - (1+t/t_0)^{-\alpha}\bigr) \quad (\alpha \ne 0), \cr
\nu \ln(1 + t/t_0) \hskip 14mm (\alpha = 0), \end{cases}
\end{equation}
with 
\begin{equation} \label{eq:nu_def}
\nu = \frac{|\Gamma|}{|\Omega|} \kappa_0 t_0 .
\end{equation}

The adiabatic analysis of Sec. \ref{sec:adiabatic} suggests five
distinct long-time regimes for the FRT density:
\begin{equation}  \label{eq:Ht_power}
H(t) \simeq \begin{cases} 
\exp\bigl(-D\lambda_0^{\infty}t + o(t)\bigr) & (\alpha < -1), \cr 
\exp\bigl(-D\lambda_0^{\kappa_\infty}t + o(t)\bigr) & (\alpha = -1), \cr 
\exp\bigl(-\tfrac{\nu}{|\alpha|} (t/t_0)^{|\alpha|} + o(t^{|\alpha|})\bigr) & (-1 < \alpha < 0), \cr
A_0 t^{-1-\nu}  &  (\alpha = 0), \cr
A_\alpha t^{-1-\alpha}  & (\alpha > 0),
\end{cases}
\end{equation}
with some unknown prefactor $A_\alpha$.  In the regimes with
exponential or stretched-exponential decay, these expressions specify
the leading behavior of the logarithm of $H(t)$; subleading
corrections affect the multiplicative prefactors and can remain
significant at moderately large times.  In the power-law decays, the
decay exponents are determined by the long-time behavior of the
reactivity.  For $\alpha<0$, the FRT distribution has finite moments
of every positive order.  At the critical value $\alpha=0$, the
exponent $\nu$ depends on the initial reactivity $\kappa_0$, the
characteristic timescale $t_0$, and the surface-to-volume ratio
$|\Gamma|/|\Omega|$.  The moment of order $n$ is finite if $0< n<
\nu$; in particular, the mean FRT diverges for $\nu \leq 1$.  For
$\alpha>0$, the integrated reactivity $\K(t)$ is finite and there is a
strictly positive probability that the particle never reacts: even
though the diffusive exploration is compact, the reactivity
$\kappa(t)$ decays too fast so that the particle may fail to react
even after an infinite number of returns to the reactive surface.
Consequently, all positive moments of the unconditional FRT are
infinite.  Note that the regime with $\alpha < -1$ was established
independently of the adiabatic approximation (see
Eq. (\ref{eq:lambda0_dominance}) and the related discussion).  In
Appendix \ref{sec:Aadiabatic}, we illustrate four asymptotic regimes
(Fig. \ref{fig:Ht_asympt}) and reveal the importance of subleading
corrections.

When the initial reactivity $\kappa(0) = \kappa_0$ is small and
$\kappa(t)$ is constant or decreasing, the four of the above
asymptotic regimes corresponding to $\alpha \geq -1$ are reduced to
the approximate relation (\ref{eq:Hhom}).  In this particular setting,
the leading-order long-time behavior of $H(t)$ does not depend on the
diffusion coefficient $D$.  For instance, one can use
Eq. (\ref{eq:lambda0_small}) to see how $D$ disappears from the
exponential function in the second regime ($\alpha = -1$).  The
transition between reaction-limited and diffusion-limited regimes
strongly depends on the shape of the confining domain, even for a
constant reactivity (see, e.g., \cite{Grebenkov13,Grebenkov18}).

Remarkably, the five asymptotic regimes in Eq. (\ref{eq:Ht_power})
coincide with those found for encounter-dependent reactivity in
Eq. (\ref{eq:He_power0}), although their prefactors and subleading
corrections generally differ.  The probabilistic reformulation
developed below clarifies the relation between these two scenarios.

\subsection{Probabilistic reformulation}
\label{sec:time_proba}

The probabilistic representation of diffusion with Robin boundary
condition is well established for space-dependent surface reactivity
\cite{Papanicolaou90}.  Motivated by this description, we consider a
more general setting when the reactivity depends both on space and
time, $\kappa(\x,t)$, and introduce the dimensionless reaction
functional
\begin{equation}  \label{eq:Lt_def}
L_t = \frac{1}{D} \int\limits_0^t \kappa(\X_{t'}, t') d\ell_{t'} .
\end{equation}
As the integral is taken with respect to the boundary local time
$\ell_t$, the reactivity is sampled at the times and positions of
encounters with the reactive surface, and each increment $d\ell_{t'}$
is weighted by the instantaneous value $\kappa(\X_{t'},t')/D$.

The first-reaction time is then defined by the stopping condition
\begin{equation}  \label{eq:tau_defL}
\tau = \inf\{ t> 0~:~ L_t > \hat{L}\}, 
\end{equation}
where $\hat{L}$ is the independent standard exponential random
variable (with mean $1$).  In particular, the survival probability
reads
\begin{equation} \label{eq:St_Lt}
S(t) = \P\{ \tau > t\} = \P\{ L_t < \hat{L}\} = \langle e^{-L_t}\rangle ,
\end{equation}
in direct analogy to Eq. (\ref{eq:Se_barLt}) for encounter-dependent
reactivity (note that a representation similar to Eq. (\ref{eq:St_Lt})
holds if the starting point is fixed).

While the encounter-dependent and time-dependent scenarios employ the
same exponential threshold $\hat{L}$ in Eqs. (\ref{eq:tau_barL},
\ref{eq:tau_defL}), their essential difference lies in the reaction
functional: in the former, $\bar{L}_t$ depends only on the terminal
boundary local time $\ell_t$, whereas in the latter, $L_t$ involves
the entire history of boundary encounters through the integral in
Eq. (\ref{eq:Lt_def}).  As a consequence, the latter generally cannot
be reduced to a prescribed function of $\ell_t$ alone that prevents a
direct application of the standard encounter-based approach.  In other
words, even though the reflected Brownian motion and its boundary
local time are determined solely by the diffusive dynamics in both
scenarios, the functional form of $L_t$ couples dynamics to
reactivity, just as in the Robin boundary value problem.  For a
constant reactivity $\kappa_0$, Eq. (\ref{eq:Lt_def}) implies $L_t =
\kappa_0 \ell_t/D = \bar{L}_t$, and the two scenarios become
identical.

The probabilistic reformulation in terms of the functional $L_t$ sheds
light onto the effect of time-dependent reactivity $\kappa(t)$.  For a
constant reactivity $\kappa_0 > 0$, the compact diffusive exploration
of a bounded domain $\Omega$ immediately implies that the reaction
occurs with probability $1$, irrespective of the value of $\kappa_0$.
In contrast, if $\kappa(t)$ decays fast enough, the particle may never
react (see Sec. \ref{sec:Ht_regimes}).  According to
Eq. (\ref{eq:tau_defL}), one has
\begin{equation}  \label{eq:tau_infty}
\P\{\tau = \infty\} = \P\{ L_\infty < \hat{L}\} = \langle e^{-L_\infty} \rangle.
\end{equation}  
Since $\langle L_\infty \rangle = \K(\infty)$, the condition
$\K(\infty) < \infty$ implies that $L_\infty$ is finite almost surely,
so that the no reaction probability is strictly positive.

\subsection{Comparison between two scenarios}

The probabilistic formulation allows us to compare the time-dependent
and encounter-dependent reaction mechanisms through their reaction
functionals defined in Eqs. (\ref{eq:barLt_def}) and
(\ref{eq:Lt_def}).  In both cases, the underlying reflected Brownian
motion and its boundary local time are identical, but the increments
of the reaction functionals are weighted differently:
\begin{equation}
d\bar{L}_t = \frac{1}{D} \kappae(\ell_t) d\ell_t, \qquad 
dL_t = \frac{1}{D} \kappa(t) d\ell_t .
\end{equation}
While equality in distribution of the two functionals at each time
would imply identical FRT laws, this cannot generally be achieved by
prescribing a deterministic encounter-dependent reactivity
$\kappae(\ell)$.  We therefore seek an approximate correspondence
between two scenarios by requiring that the instantaneous reactivities
sampled along reflected trajectories coincide {\it on average}:
\begin{equation}  \label{eq:equivalence_general}
\langle \kappae(\ell_t) \rangle = \kappa(t) \qquad (t \geq 0).
\end{equation}
For a given time-dependent reactivity $\kappa(t)$, one therefore aims
at finding an encounter-dependent reactivity $\kappae(\ell)$
satisfying Eq. (\ref{eq:equivalence_general}).  This is a challenging
inverse problem that requires the knowledge of the boundary local time
distribution in a given domain
\cite{Grebenkov19,Grebenkov20c,Grebenkov22b}.
Moreover, the matching condition (\ref{eq:equivalence_general}) does
not necessarily guarantee the equality of the associated FRTs $\tau$
and $\taue$ in distribution.

We therefore adopt a simpler mean-field prescription, replacing the
average of the encounter-dependent reactivity by its value at the mean
boundary local time:
\begin{equation}  \label{eq:equivalence_mean}
\kappae(\langle \ell_t\rangle) = \kappa(t) \qquad (t \geq 0).
\end{equation}
Although this replacement is heuristic, the mean boundary local time
itself is known exactly for a uniform initial distribution (see
Appendix \ref{sec:Aencounter} and Ref. \cite{Grebenkov19}),
\begin{equation}  \label{eq:mean_BLT}
\langle \ell_t \rangle = v t ,
\end{equation}
with $v$ from Eq. (\ref{eq:v_def}).  This leads to the explicit
mapping
\begin{equation}  \label{eq:equivalence}
\kappae(\ell) = \kappa(\ell/v)  \qquad \Leftrightarrow \qquad  \Ke(\ell) = \K(\ell/v) , 
\end{equation}
where $\K(t)$ is given by Eq. (\ref{eq:Kint}).  This relation provides
a heuristic mapping from the original problem with a given $\kappa(t)$
to an encounter-dependent description formulated in terms of a random
threshold $\hat{\ell}$, with the distribution
\begin{equation}  \label{eq:Psi_eff}
\Psi(\ell) = e^{- \K(\ell/v)}.
\end{equation}
As a consequence, Eq. (\ref{eq:Ht_Ut}) reads
\begin{equation}  \label{eq:EBA}
\Heba(t) = \int\limits_0^\infty d\ell \, \frac{\kappa(\ell/v)}{D} e^{-\K(\ell/v)}  \, U(\ell,t).
\end{equation}
Equation (\ref{eq:EBA}) gives the exact FRT density of the
encounter-dependent model defined by Eq. (\ref{eq:equivalence}).  If
the mapping (\ref{eq:equivalence}) captures correctly the similarities
between time-dependent and encounter-dependent scenarios,
Eq. (\ref{eq:EBA}) may serve an analytical approximation of the FRT
density of the original time-dependent problem.  In the following, we
compare the PDFs $H(t)$ and $\Heba(t)$ to qualify the accuracy of the
mapping, as well as the range of its applicability.

The heuristic equivalence qualitatively explains why the long-time
asymptotic decay (\ref{eq:Ht_power}) for time-dependent reactivity
$\kappa(t)$ exhibits the same five regimes as found in
Eq. (\ref{eq:He_power0}) for encounter-dependent reactivity
$\kappae(\ell)$.  Once the functions $\kappa(t)$ and $\kappae(\ell)$
are matched via Eq. (\ref{eq:equivalence}), one can translate one
relation to the other.  For instance, the power-law model
(\ref{eq:kappa_model}) for time-dependent reactivity is translated
into the equivalent model
\begin{equation}  \label{eq:kappa_ell} 
\kappae(\ell) = \kappa_0 (1 + \ell/\ell_0)^{-1-\alpha} ,
\end{equation}
where
\begin{equation}  \label{eq:ell0}
\ell_0 = vt_0 = \frac{|\Gamma|}{|\Omega|} Dt_0 .
\end{equation}
This model corresponds to the random threshold $\hat{\ell}$, whose
distribution is given by Eq. (\ref{eq:Psi_kappa}) as
\begin{equation}  \label{eq:Psi_model}
\Psi(\ell) = \begin{cases} \exp\bigl(- \frac{\nu}{\alpha} \bigl(1 - (1 + \ell/\ell_0)^{-\alpha}\bigr) \bigr) \quad (\alpha \ne 0), \cr
(1 + \ell/\ell_0)^{-\nu}  \hskip 27mm (\alpha = 0), \end{cases}
\end{equation}
and
\begin{equation}  \label{eq:psi_model}
\psi(\ell) = \frac{\nu}{\ell_0} (1+\ell/\ell_0)^{-1-\alpha} \Psi(\ell).
\end{equation}
In particular, one sees that $\Psi(\infty) = 0$ for $\alpha \leq 0$,
whereas $\Psi(\infty) = e^{-\nu/\alpha}$ for $\alpha > 0$.

At the same time, the heuristic equivalence does not necessarily imply
quantitative agreement.  For instance, when $\alpha > 0$, the
probability of no reaction in the encounter-dependent setting was
given explicitly by Eq. (\ref{eq:Pe_infty}).  In turn, the no reaction
probability for the time-dependent scenario is determined by
Eq. (\ref{eq:tau_infty}), which is generally not reducible to
$e^{-\K(\infty)}$.  In fact, the Jensen inequality for the convex
function $e^{-x}$ implies
\begin{equation}
\P\{ \tau = \infty\} = \langle e^{- L_\infty}\rangle \geq e^{-\langle L_\infty\rangle} = e^{-\Ke(\infty)} 
= \P\{ \taue = \infty\}.
\end{equation}
In other words, the mapping recovers the condition $K(\infty)<\infty$
for a positive probability of never reacting, but the actual value of
this probability is underestimated.

\section{Numerical simulations}
\label{sec:numerics}

To illustrate similarities and differences between the two scenarios
of reaction kinetics, we study diffusion (with $D = 1$) in a
spherical shell $\Omega = \{ \x\in\R^3 ~:~ R < |\x|<L\}$, with a
reactive inner sphere of radius $R = 1$ and a reflecting outer sphere
of radius $L = 3$ (Fig. \ref{fig:scheme}(b)).  The PDF $H(t)$ is
computed by Monte Carlo simulations based on a walk-on-spheres
algorithm with boundary reflections (Appendix \ref{sec:direct}),
validated against exact results for constant reactivity
\cite{Grebenkov18,Grebenkov20b} and independent finite-difference
calculations (Appendix \ref{sec:FDM}).  The rotational symmetry of
$\Omega$ facilitates the computation of $\Heba(t)$ from
Eq. (\ref{eq:EBA}) by using the Laplace-domain expression for
$U(\ell,t)$, as described in Appendix \ref{sec:shell}.  Despite the
focus on the spherical geometry in this section, the theoretical
results obtained in Secs. \ref{sec:encounter} and \ref{sec:time} are
valid for general bounded domains with smooth enough boundaries.

\begin{figure}
\centering
\includegraphics[width=0.49\columnwidth]{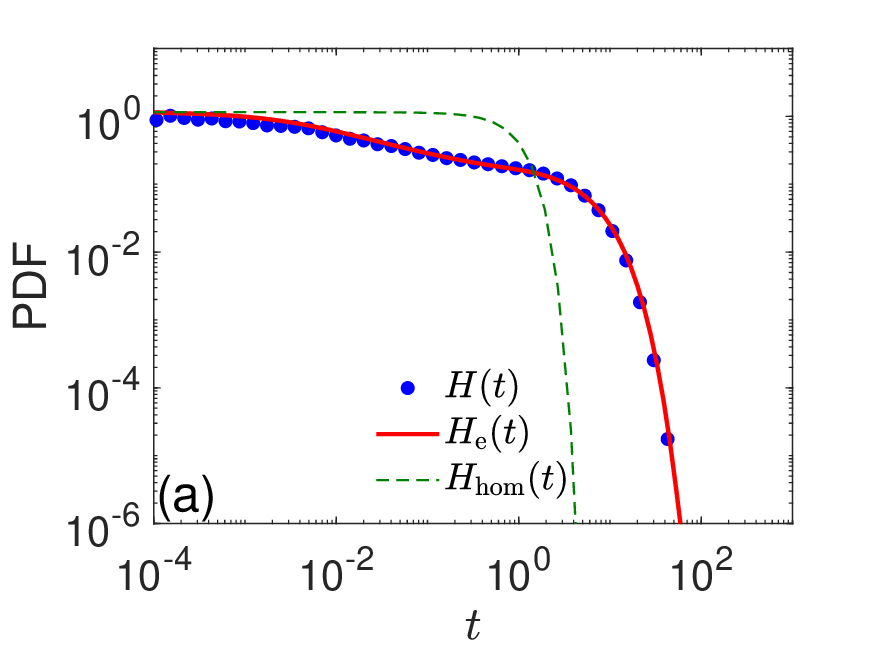} 
\includegraphics[width=0.49\columnwidth]{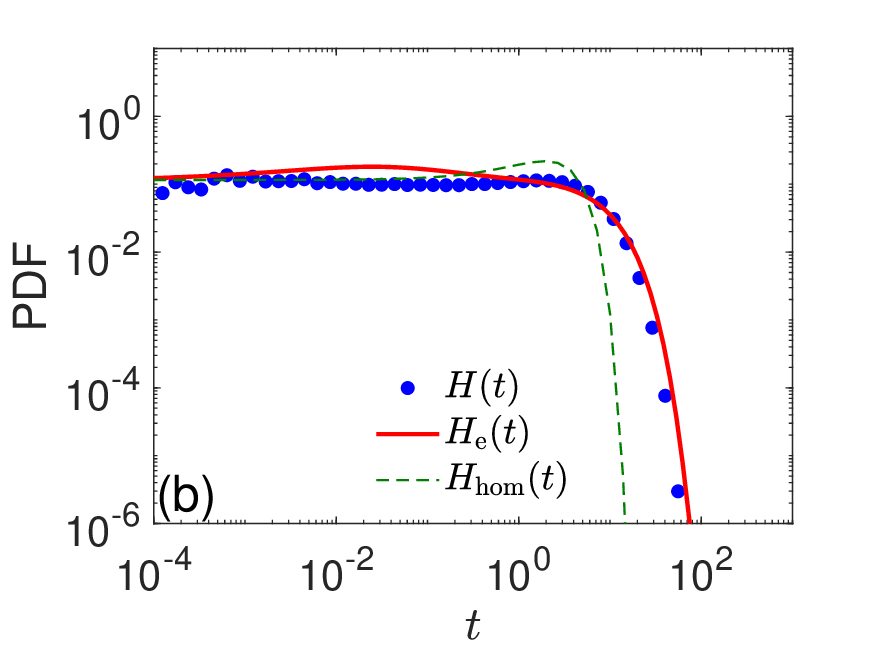} 
\includegraphics[width=0.49\columnwidth]{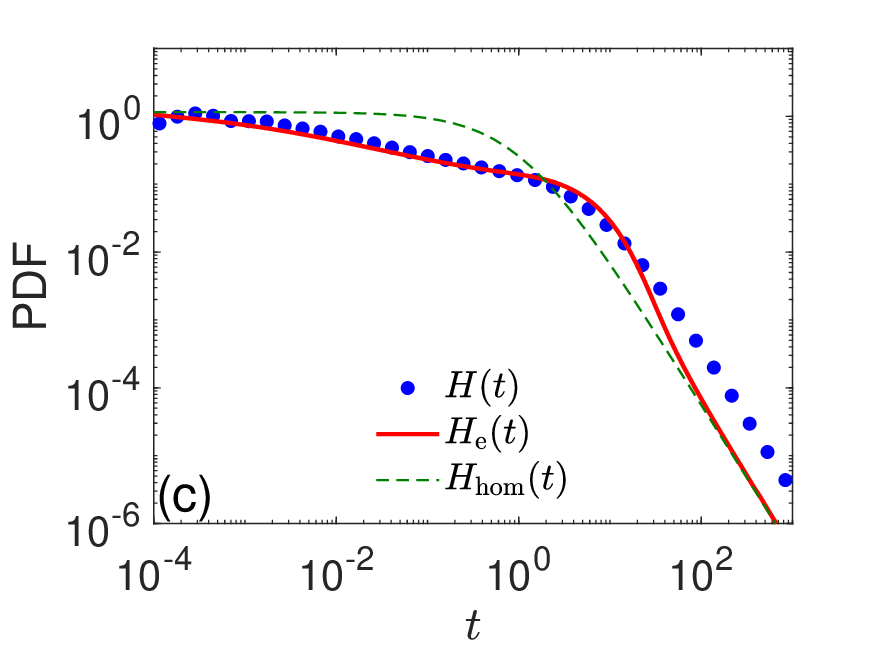} 
\includegraphics[width=0.49\columnwidth]{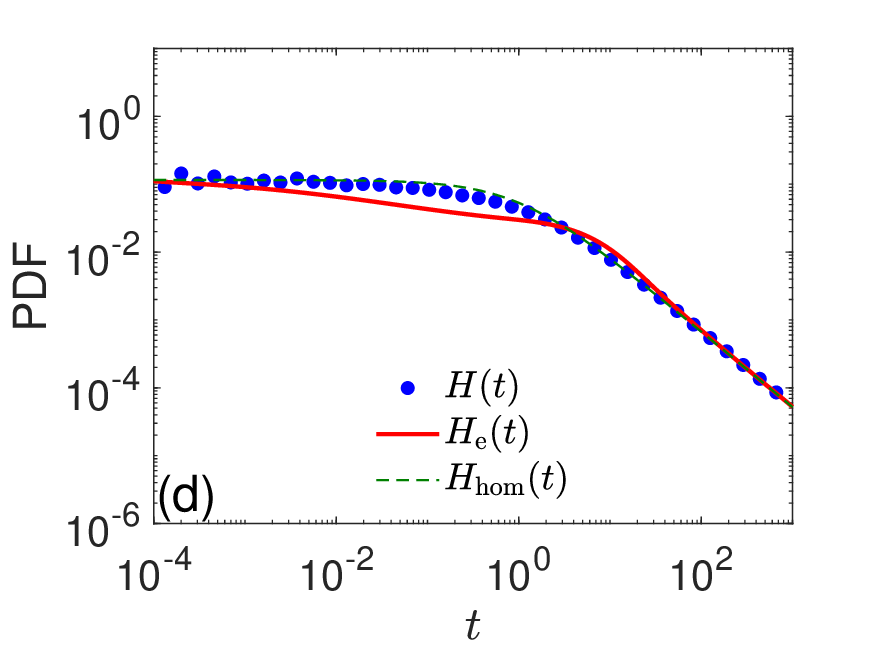} 
\includegraphics[width=0.49\columnwidth]{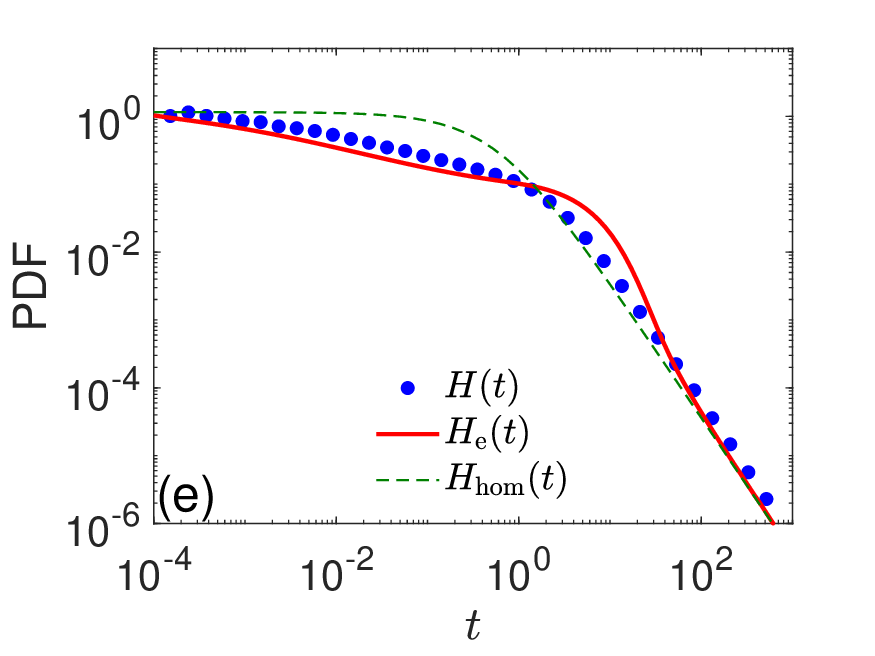} 
\includegraphics[width=0.49\columnwidth]{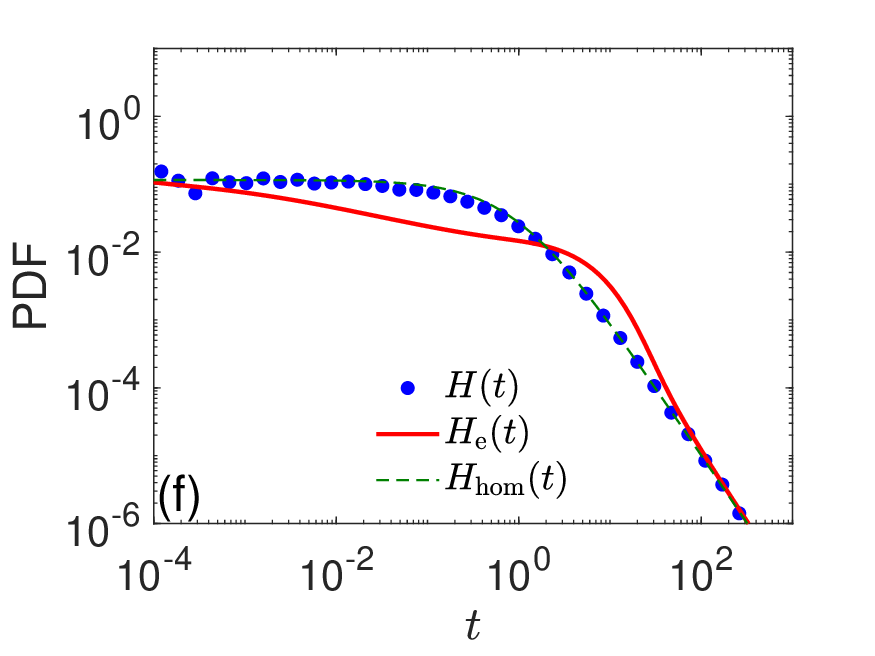} 
\caption{
Volume-averaged PDF of the first-reaction time on a spherical target
of radius $R$, surrounded by a reflecting concentric sphere of radius
$L$, with power-law reactivity model (\ref{eq:kappa_model}).  Circles
show $H(t)$, solid line presents $\Heba(t)$ from Eq. (\ref{eq:EBA}),
whereas dashed line indicates the low-reactivity approximation
(\ref{eq:Hhom}).  We used $t_0 = 1$, $D = 1$, $R = 1$, $L = 3$, two
values $\kappa_0 = 10$ (left column) and $\kappa_0 = 1$ (right
column), and three values of $\alpha$: $-2$ (top row), $0$ (middle
row), and $1$ (bottom row).  }
\label{fig:target1}
\end{figure}

We consider the power-law reactivity model (\ref{eq:kappa_model}),
with $t_0=1$, two initial reactivities $\kappa_0 =10$ and $\kappa_0 =
1$, and three exponents $\alpha=-2,0,1$ that model increasing,
critically decreasing, and rapidly decreasing reactivities
$\kappa(t)$.  Figure \ref{fig:target1} compares the PDFs $H(t)$ and
$\Heba(t)$, thus testing the adequacy of the mapping
(\ref{eq:equivalence}) between the two scenarios.  The low-reactivity
approximation (\ref{eq:Hhom}) is also included for comparison.

When $\alpha = -2$, the reactivity $\kappa(t)$ grows with time, and
$\Heba(t)$ accurately reproduces $H(t)$ (top row).  The excellent
agreement for $\kappa_0 = 10$ [panel (a)] is slightly deteriorated for
$\kappa_0 = 1$ [panel (b)].  In this case, both time-dependent and
encounter-dependent scenarios result in almost identical FRT
distributions, once their reactivities are matched according to
Eq. (\ref{eq:equivalence}).  The improved agreement at high reactivity
is expected because both models approach the perfectly absorbing limit
as $\kappa_0\to\infty$, in which the reaction occurs upon the first
encounter with the reactive surface, independently of the subsequent
evolution of reactivity.  Expectedly, the low-reactivity approximation
fails in this regime, especially for $\kappa_0 = 10$.

For $\alpha = 0$ (middle row), the reactivity $\kappa(t)$ decreases in
the critical regime, as $1/t$.  For $\kappa_0 = 10$ [panel (c)], we
have $\nu \approx 1.15 > 1$ so that the mean FRT is finite, whereas
panel (d) for $\kappa_0 = 1$ corresponds to $\nu \approx 0.115 < 1$
with an infinite mean FRT.  The associated PDF $H(t)$ looks quite
different in the two panels, being much broader on the panel (d).  On
panel (c), $H(t)$ and $\Heba(t)$ are almost identical at short times
but deviations are observed at large $t$: even though both $H(t)$ and
$\He(t)$ exhibit the same power-law decay, the prefactors are
different.  In turn, the agreement is better on panel (d).  The
low-reactivity approximation still fails for $\kappa_0 = 10$, but
becomes good for $\kappa_0 = 1$.

Finally, in the case $\alpha = 1$ (bottom row), the reactivity
$\kappa(t)$ decreases as $t^{-2}$, implying that some particles never
react.  Since some particles never react, the density of finite FRT
integrates to $1-S(\infty)<1$.  At long times, this density decays as
$H(t)\propto t^{-2}$, as predicted by Eq. (\ref{eq:Ht_power}).  The
same decay $\He(t)\propto t^{-2}$ is observed, but the prefactor is
slightly different.  In turn, the low-reactivity approximation turns
out to be very accurate on panel (f).

\section{Conclusions}
\label{sec:conclusion}

We investigated diffusion-controlled reactions with time-dependent
surface reactivity, for which the conventional diffusion equation with
a time-dependent Robin boundary condition generally precludes direct
application of standard Laplace transforms and spectral methods.  We
developed complementary spectral and probabilistic approaches to
characterize the FRT statistics.

An adiabatic approximation based on instantaneous Robin eigenmodes
revealed five long-time regimes of the FRT distribution, exhibiting
exponential, stretched-exponential, or power-law decay.  The critical
reactivity $\kappa(t)\propto 1/t$ yields $H(t)\propto t^{-1-\nu}$,
with an exponent $\nu$ that depends on the reaction parameters, so
that the mean FRT may diverge even in a bounded domain.  For faster
power-law decay of $\kappa(t)$, the probability of never reacting
becomes strictly positive.

We also established a probabilistic formulation of time-dependent
reactivity in terms of reflected Brownian motion and boundary local
time.  The reaction event is determined by the reaction functional
$L_t$, which weights boundary encounters according to the reactivity
at their occurrence times.  Unlike its encounter-dependent counterpart
$\bar{L}_t$, which depends only on the terminal boundary local time
$\ell_t$, $L_t$ depends on the entire history of encounters.  This
distinction clarifies the relation between the two reaction
mechanisms.

Relying on this relation, we proposed a heuristic mapping
(\ref{eq:equivalence}) between the two scenarios that relates the
reactivities $\kappa(t)$ and $\kappae(\ell)$.  This mapping reproduces
the five leading asymptotic regimes, although prefactors and
subleading corrections generally differ.  Moreover, we established
that the probability of never reacting in the time-dependent model is
bounded from below by that of the corresponding encounter-dependent
model.
We assessed the accuracy of the mapping by comparing the PDFs $H(t)$
and $\Heba(t)$, evaluated by different independently validated
numerical methods.  For the considered power-law reactivity model, the
two matched scenarios gave very close PDFs for an increasing
reactivity ($\alpha = -2$) but revealed deviations for decreasing
reactivities ($\alpha = 0$ and $\alpha = 1$).  This observation
motivates further investigation of more refined mappings that account
for fluctuations of the boundary local time, and extending the
analysis to spatially heterogeneous surfaces with prescribed space-
and time-dependent reactivities.

Finally, the framework also applies to diffusion with time-dependent
diffusivity $D(t)$ and partial reactivity
\cite{deSwiet96,Valiullin01,Qian06,Molini11}.  Indeed, the time change
$T=\int\nolimits_0^t dt' D(t')/D$ transforms this problem into
diffusion with constant diffusivity $D$ and an effective
time-dependent reactivity $\kappa_{\rm eff}(T) = D\kappa_0/D(t(T))$.
This correspondence provides another application of the present
approach to diffusive transport with time-varying properties.

\begin{acknowledgments}
The authors used OpenAI's ChatGPT (GPT-5.6) during the preparation of
this manuscript to assist with language editing and verification of
analytical derivations.  The authors assume full responsibility for
all scientific content.
\end{acknowledgments}

\section*{Data availability}

The data that support the findings of this study are available from
the corresponding author upon reasonable request.

\appendix
\section{Encounter-based approach}
\label{sec:Aencounter}

Most properties of diffusion-controlled reactions, including the
statistics of the boundary local time $\ell_t$, can be represented in
terms of eigenvalues $\mu_k^{(p)}$ and eigenfunctions $V_k^{(p)}$ of
the Steklov problem for the modified Helmholtz equation with a
parameter $p \geq 0$ \cite{Grebenkov19,Grebenkov20}:
\begin{equation}  \label{eq:Helm}
(p - D\Delta) V_k^{(p)} = 0  \quad \textrm{in}~\Omega, \quad 
\begin{cases} \partial_n V_k^{(p)} = \mu_k^{(p)} V_k^{(p)} \quad \textrm{on}~\Gamma, \cr  
\partial_n V_k^{(p)} = 0 \hskip 10mm \textrm{on}~\pa_0. \end{cases}
\end{equation}
When $\Gamma$ is bounded, the spectrum is discrete, and the
eigenvalues $\mu_k^{(p)}$ can be enumerated by the index $k =
0,1,2,\ldots$ in an increasing order \cite{Levitin,Grebenkov27}.  The
restrictions of $V_k^{(p)}$ onto $\Gamma$ are the eigenfunctions of
the Dirichlet-to-Neumann operator $\M_p$ associated to the boundary
value problem (\ref{eq:Helm}).  These restrictions are orthogonal to
each other and form a complete basis in the space $L^2(\Gamma)$ of
square-integrable functions on $\Gamma$.  We fix their normalization
by imposing that
\begin{equation}  \label{eq:vk_norm}
\int\limits_\Gamma |V_k^{(p)}|^2 = 1.
\end{equation}
When $\Omega$ is connected (as we always assume), the principal
eigenvalue $\mu_0^{(p)}$ is simple. 
Integrating the modified Helmholtz equation in Eq. (\ref{eq:Helm})
over $\Omega$ and applying the divergence theorem yield the identity
\begin{equation}  \label{eq:identity1}
\frac{p}{D} \int\limits_{\Omega} d\x \, V_k^{(p)} = \mu_k^{(p)} \int\limits_{\Gamma} d\x \, V_k^{(p)} .
\end{equation}

\subsection{Mean boundary local time}

For a fixed starting point $\x_0$, the mean boundary local time can be
written directly as
\begin{equation}
\langle \ell_t \rangle_{\x_0} = D \int\limits_0^t dt' \int\limits_{\Gamma} d\x \, G_0(\x,t'|\x_0,0),
\end{equation}
where $G_0(\x,t'|\x_0,0)$ is the heat kernel with Neumann boundary
condition (see, e.g., \cite{Grebenkov19,Grebenkov22b}).  The integral
of this expression over $\x_0 \in \Omega$ yields immediately
Eq. (\ref{eq:mean_BLT}) due to the symmetry and normalization of this
kernel.  An alternative derivation can be achieved by using spectral
representations in terms of Steklov eigenfunctions \cite{Grebenkov19}.

\subsection{PDF of the first-crossing time}

As the boundary local time $\ell_t$ is a nondecreasing process, one
can relate its statistics to that of the first-crossing time $\T_\ell$
of a fixed threshold $\ell$:
\begin{equation}
\T_\ell = \inf\{ t>0~:~\ell_t > \ell\}.  
\end{equation}
The spectral expansion for the moment-generating function of this
random variable was derived in \cite{Grebenkov20}.  When the starting
point is uniformly distributed, one gets (see Eq. (S103) in the SM of
\cite{Grebenkov20}):
\begin{align}  \nonumber
\tilde{U}(\ell,p) & = \langle e^{-p\T_\ell} \rangle = \int\limits_0^\infty dt \, e^{-pt} \, U(\ell,t) \\  \label{eq:Up}
& = \frac{D}{p|\Omega|} \sum\limits_{k=0}^\infty \mu_k^{(p)} e^{-\mu_k^{(p)}\ell} 
\biggl|\int\limits_{\Gamma} V_k^{(p)}\biggr|^2 
\end{align}
(here and below, tilde denotes Laplace-transformed quantities with
respect to time $t$).  Since $V_k^{(p)}|_{\Gamma}$ are the
eigenfunctions of the Dirichlet-to-Neumann operator $\M_p$, the above
spectral expansion can also be written as
\begin{equation}  \label{eq:Up2}
\tilde{U}(\ell,p) = \frac{D}{p|\Omega|}  \bigl(1, \M_p e^{-\M_p\ell} 1\bigr)_{L^2(\Gamma)}.
\end{equation}

The moment-generating function $\tilde{U}(\ell,p)$ plays the central
role in studying first-reaction times for various surface reaction
mechanisms.  We recall that the FRT $\taue$ was introduced in
Eq. (\ref{eq:tau_def}) as the first instance when the boundary local
time $\ell_t$ exceeds a random threshold $\hat{\ell}$, with a given
distribution $\Psi(\ell) = \P\{\hat{\ell} > \ell\}$.  One has thus
$\taue = \T_{\hat{\ell}}$, so that the moment-generating function of
$\taue$ is the average of $\tilde{U}(\hat{\ell},p)$ over $\hat{\ell}$:
\begin{equation}  \label{eq:Hp_Up}
\tilde{\He}(p) = \langle e^{-p\taue}\rangle = \langle e^{-p\T_{\hat{\ell}}}  \rangle 
= \int\limits_0^\infty d\ell \, \psi(\ell) \, \tilde{U}(\ell,p) \,,
\end{equation}
where $\psi(\ell) = - \partial_\ell \Psi(\ell)$ is the PDF of the
threshold $\hat{\ell}$.  This relation remains valid even when the
threshold distribution is defective, with a nonzero probability
$\Psi(\infty)$ of an infinite threshold.  In this case, the FRT is
infinite with probability $\Psi(\infty)$, whereas its density $\He(t)$
describes only finite reaction times and integrates to
$1-\Psi(\infty)$.  Substituting Eq. (\ref{eq:Up}) into
Eq. (\ref{eq:Hp_Up}) gives
\begin{align}  \label{eq:Hp_spectral}
\tilde{\He}(p) & = \frac{D}{p|\Omega|} \sum\limits_{k=0}^\infty \mu_k^{(p)} \, \Upsilon(\mu_k^{(p)}) 
\biggl|\int\limits_{\Gamma} V_k^{(p)}\biggr|^2 \\  \label{eq:Hp_spectral_Mp}
& = \frac{D}{p|\Omega|} \bigl(1, \M_p \, \Upsilon(\M_p) 1\bigr)_{L^2(\Gamma)} ,
\end{align}
where
\begin{equation}  \label{eq:Upsilon}
\Upsilon(\mu) = \langle e^{-\mu \hat{\ell}} \rangle = \int\limits_0^\infty d\ell \, e^{-\mu \ell} \psi(\ell)
\end{equation}
is the moment-generating function of the threshold $\hat{\ell}$ (Table
I from the SM of \cite{Grebenkov20} reports several examples).

The PDF of the first-crossing time can be found via the inverse
Laplace transform of the moment-generating function with respect to
$p$:
\begin{equation}  \label{eq:Ut_inverse}
U(\ell,t) = \L^{-1} \bigl\{ \tilde{U}(\ell,p)  \bigr\}.
\end{equation}
Likewise, the inverse Laplace transform of Eq. (\ref{eq:Hp_Up}) yields
Eq. (\ref{eq:Ht_Ut}) for the PDF of the FRT $\taue$.  Similar
relations hold for all the moments whenever they exist; for instance,
if the mean FRT is finite, one has
\begin{equation}  \label{eq:mean_FRT}
\langle \taue\rangle = \int\limits_0^\infty dt \, t\, \He(t)
= \int\limits_0^\infty d\ell \, \psi(\ell) \, \langle \T_\ell\rangle \,,
\end{equation}
where $\langle \T_\ell\rangle$ is the mean first-crossing time (with
the uniform starting point).

\subsection{Mean first-crossing time}

In order to evaluate the mean first-crossing time, we consider the
generating function $\tilde{U}(\ell,p)$ in the limit of small $p$.
For this purpose, we recall that (see \cite{Grebenkov19c})
\begin{subequations}  \label{eq:Steklov_p0}
\begin{align}  
\mu_0^{(p)} & = \frac{p}{v} \bigl(1 - b_0 p + O(p^2)\bigr) \to 0,   \\
\mu_k^{(p)} & \to \mu_k^{(0)} > 0  \qquad (k > 0), \\
V_k^{(p)} & = V_k^{(0)} + u_k\, p + O(p^2) \qquad (k \geq 0),
\end{align}
\end{subequations}
with $v$ given by Eq. (\ref{eq:v_def}).  Treating separately the terms
$k = 0$ and $k > 0$ in Eq. (\ref{eq:Up}), one gets in the leading
order
\begin{align} \nonumber
\tilde{U}(\ell,p) & = (1 - b_0 p) \bigl(1 - p\ell/v\bigr) \\
\label{eq:auxil2}
& + p \frac{D}{|\Omega|} \sum\limits_{k=1}^\infty \mu_k^{(0)} e^{-\mu_k^{(0)}\ell} \biggl|\int\limits_{\Gamma} u_k\biggr|^2 + o(p), 
\end{align}
where we used that $V_0^{(0)} = 1/\sqrt{|\Gamma|}$ and
$\int\nolimits_{\Gamma} u_0 = 0$.  In the limit $p\to 0$, the identity
(\ref{eq:identity1}) yields for $k > 0$,
\begin{equation}
\int\limits_{\Gamma} u_k = \frac{1}{D\mu_k^{(0)}}\int\limits_{\Omega} V_k^{(0)}  ,
\end{equation}
where we used that $V_k^{(0)}|_{\Gamma}$ is orthogonal to a constant
function $V_0^{(0)}$.  Substituting this expression into
Eq. (\ref{eq:auxil2}), we deduce the mean first-crossing time:
\begin{align} \nonumber
\langle \T_\ell\rangle & = \int\limits_0^\infty dt \, t \, U(\ell,t) 
= - \lim\limits_{p\to 0} \partial_p \tilde{U}(\ell,p) \\     \label{eq:Tell_mean}
& = \ell/v  + b_0 
- \frac{1}{D|\Omega|} \sum\limits_{k=1}^\infty \frac{e^{-\mu_k^{(0)}\ell}}{\mu_k^{(0)}} \biggl|\int\limits_{\Omega} V_k^{(0)}\biggr|^2 .
\end{align}
Setting $\ell = 0$, one gets the spectral expansion for the
volume-averaged mean first-passage time to $\Gamma$:
\begin{equation}   \label{eq:T0_mean}
\langle \T_0\rangle = b_0 
- \frac{1}{D|\Omega|} \sum\limits_{k=1}^\infty \frac{1}{\mu_k^{(0)}} \biggl|\int\limits_{\Omega} V_k^{(0)}\biggr|^2 \leq b_0.
\end{equation}
A formal expression for $b_0$ was derived in \cite{Grebenkov19} (see
also \cite{Grebenkov25})
\begin{equation} \label{eq:b0_G}
b_0 = \frac{|\Omega|}{D|\Gamma|^2} \int\limits_{\Gamma} d\x \int\limits_{\Gamma} d\x_0 \, {\mathcal G}(\x,\x_0),
\end{equation}
where ${\mathcal G}(\x,\x_0)$ is the Neumann Green's function.  For
instance, one has
\begin{equation}  \label{eq:b0_shell}
b_0 = \frac{(L-R)^3(5L^3+6L^2R+3LR^2+R^3)}{15RD (L^3-R^3)} 
\end{equation}
for a spherical target of radius $R$, surrounded by a concentric
reflecting sphere of radius $L$.  In fact, the rotational symmetry of
this domain cancels the sum in Eq. (\ref{eq:T0_mean}), implying $b_0 =
\langle \T_0\rangle$.

Subtracting Eqs. (\ref{eq:Tell_mean}, \ref{eq:T0_mean}), one gets rid
of $b_0$ as
\begin{equation}  \label{eq:Tell_mean2}
\langle \T_\ell \rangle = \langle \T_0\rangle + \ell/v 
+ \sum\limits_{k=1}^\infty \frac{1-e^{-\mu_k^{(0)}\ell}}{D|\Omega| \mu_k^{(0)}} \biggl|\int\limits_{\Omega} V_k^{(0)}\biggr|^2  .
\end{equation}
We see that $\langle\T_\ell\rangle$ is a strictly increasing function
of $\ell$, ranging from $\langle \T_0\rangle$ to infinity, with the
linear scaling $\langle T_\ell\rangle = \ell/v +b_0 + o(1)$ at large
$\ell$, where $v$ is given by Eq. (\ref{eq:v_def}).  In addition,
Eq. (\ref{eq:Tell_mean2}) implies a simple bound: $\langle \T_\ell
\rangle \geq \langle \T_0\rangle + \ell/v$.

According to Eq. (\ref{eq:mean_FRT}), the mean FRT for a general
surface reaction is determined by the probability density $\psi(\ell)$
of the threshold $\hat{\ell}$ as
\begin{equation}  \label{eq:meanFRT_general}
\langle \taue\rangle = \langle \T_0\rangle + \langle \hat{\ell}\rangle/v +  
\sum\limits_{k=1}^\infty \frac{1-\Upsilon(\mu_k^{(0)})}{D|\Omega| \mu_k^{(0)}} \biggl|\int\limits_{\Omega} V_k^{(0)}\biggr|^2 ,
\end{equation}
where $\Upsilon(\mu)$ was defined in Eq. (\ref{eq:Upsilon}).  In
particular, if $\Psi(\infty)=0$, then
\begin{equation}
\langle \hat{\ell} \rangle = \int\limits_0^\infty d\ell \, \ell \, \psi(\ell) 
= \int\limits_0^\infty d\ell \, \Psi(\ell) = - \lim\limits_{\mu\to 0} \partial_\mu \Upsilon(\mu).
\end{equation}
In turn, if $\Psi(\infty) > 0$, the unconditional mean $\langle
\hat{\ell}\rangle$ is infinite, implying $\langle \taue\rangle
= \infty$.  In Appendix \ref{sec:power-law}, we provide an explicit
form of $\Upsilon(\mu)$ and discuss its properties for the power-law
reactivity model.

\subsection{Asymptotic behavior of $U(\ell,t)$}

Since the PDF $U(\ell,t)$ plays the central role in the
encounter-based scenario, it is instructive to briefly discuss its
asymptotic behavior at small and large $t$.

(i) At long times, the boundary local time $\ell_t$ gets concentrated
around its mean value $\langle \ell_t\rangle = vt$, with the variance
behaving linearly with $t$ at long times (see Eq. (2.46) from
\cite{Grebenkov25}, as well as \cite{Grebenkov07}): $\sigma_t^2
\approx 2 v^2 b_0 t + O(1)$, where $b_0 > 0$ is given by
Eq. (\ref{eq:b0_G}) for generic domain and by Eq. (\ref{eq:b0_shell})
for the spherical shell.
In the joint limit of large $t$ and $\ell$, with
$\ell-vt=O(\sqrt{t})$, the Gaussian approximation for boundary local
time yields the following leading-order approximation for the PDF of
the first-crossing time for $\ell > 0$:
\begin{align} \nonumber
U(\ell,t) & = -\partial_t \P\{ \T_\ell > t\} = -\partial_t \P\{ \ell_t \leq \ell\} \\  \nonumber
& \approx - \partial_t \int\limits_0^\ell d\ell' \, \frac{e^{-(\ell' - vt)^2/(2\sigma_t^2)}}{\sqrt{2\pi \sigma_t^2}} \\  \label{eq:U_Gauss}
& \approx \frac{e^{-(\ell - vt)^2/(4b_0 v^2 t)}}{\sqrt{4\pi b_0 t}} \,. 
\end{align}
As a central limit result, this expression is only valid near its
maximum, which is located at $t_{\rm max} = \sqrt{b_0^2 + \ell^2/v^2}
- b_0 \approx \ell/v - b_0 + o(1)$ at large $\ell$.  Even though this
relation is not applicable at short times, the integral of the
right-hand side over $t$ from $0$ to $\infty$ is equal to $1$,
ensuring the correct normalization as an approximate PDF.

(ii) At short times, one can analyze the asymptotic behavior of
Eq. (\ref{eq:Up2}) as $p\to \infty$.  When $\Gamma = \pa$ is smooth,
each eigenvalue behaves as
\begin{equation}  \label{eq:muk_pinf}
\mu_k^{(p)} \approx \sqrt{p/D} - \frac{d-1}{2} \, H_{\rm max}  + O(1/\sqrt{p}) \qquad (p\to \infty), 
\end{equation}
where $H_{\rm max}$ is the maximum value of the mean curvature $H(\x)$
on $\Gamma$.  Here $H(\x)$ is the average of the principal curvatures,
with the outward normal to $\Omega$ and the sign convention $H = -1/R$
on an inner sphere of radius $R$ (see page 28 of \cite{Grebenkov27},
as well as the original result for Robin Laplacian in
\cite{Pankrashkin15}).  Assuming that this asymptotic relation remains
valid in the presence of reflecting boundary $\pa_0$ (i.e., when
$\Gamma \ne \pa$), its substitution to Eq. (\ref{eq:Up2}) and the
inverse Laplace transform yield in the leading order
\begin{equation}  \label{eq:Ut_asympt_t0}
U(\ell,t) \approx \frac{v}{\sqrt{\pi Dt}}\, e^{- \ell^2/(4Dt) + \tfrac12(d-1)\ell H_{\rm max}}  \quad (t\to 0).
\end{equation}
For instance, when $\Omega$ is a spherical shell and $\Gamma$ is the
inner sphere of radius $R$, the mean curvature is $-1/R$, yielding
$H_{\rm max} = -1/R$.  We emphasize that Eq. (\ref{eq:Ut_asympt_t0})
remains formal and conjectural; in particular, even though the
asymptotic relation (\ref{eq:muk_pinf}) is valid for each $k$, its
applicability in the spectral expansion (\ref{eq:Up2}), including the
terms with $k\to \infty$, has to be justified.

\section{Power-law model of reactivity}
\label{sec:power-law}

We analyze here the power-law model (\ref{eq:kappa_ell}) of the
encounter-dependent reactivity that was inherited from the model
(\ref{eq:kappa_model}) in the time-dependent scenario.  This model is
realized by choosing the random threshold $\hat{\ell}$ with the
probability law (\ref{eq:Psi_model}).  After substitution of this law
into Eq. (\ref{eq:Upsilon}), we need to evaluate
\begin{equation}  \label{eq:Upsilon_power}
\Upsilon(\mu) 
= \nu \int\limits_0^\infty dx \, e^{-x \mu \ell_0} \, (1 + x)^{-\alpha-1} e^{-\nu (1 - (1+x)^{-\alpha})/\alpha} ,
\end{equation}
where $\nu$ was given in Eq. (\ref{eq:nu_def}).  We distinguish three
cases:

(i) For $\alpha < 0$, one can change the integration variable $z =
(1+x)^{-\alpha}$ to get
\begin{equation}  \label{eq:Upsilon_negative}
\Upsilon(\mu) = \frac{\nu}{|\alpha|} e^{\nu/|\alpha| + \mu \ell_0} 
\int\limits_1^\infty dz \, e^{-(\nu/|\alpha|)z - (\mu \ell_0) z^{1/|\alpha|}} .
\end{equation}
Expanding the second exponential function into a Taylor series, one
gets
\begin{equation}
\Upsilon(\mu) = e^{\nu/|\alpha| + \mu \ell_0} \sum\limits_{n=0}^\infty \frac{(-\mu \ell_0)^n}{n!\, (\nu/|\alpha|)^{n/|\alpha|}}  \,
\Gamma\biggl(1 + \frac{n}{|\alpha|}, \frac{\nu}{|\alpha|}\biggr),
\end{equation}
where $\Gamma(\alpha, z) = \int\nolimits_z^\infty dx \, x^{\alpha-1}
e^{-x}$ is the upper incomplete Gamma function.  
Depending on the value of $|\alpha|$, the radius of convergence of
this formal series is either infinite ($\alpha < - 1$), or finite
($\alpha = -1$), or zero ($-1 < \alpha < 0$).  In the latter case, it
is an asymptotic expansion as $\mu\to 0^+$, not a convergent Taylor
series, whereas $\Upsilon(\mu)$ is not analytic at $0$ (see Appendix
\ref{sec:Along}).

For $\alpha = -0.5$, the integral (\ref{eq:Upsilon_negative}) is
reduced to
\begin{equation}
\Upsilon(\mu) = \nu \sqrt{\pi /(\mu \ell_0)} \,  \erfcx\bigl(\sqrt{\mu \ell_0} + \nu/\sqrt{\mu \ell_0}\bigr),
\end{equation}
where $\erfcx(z) = e^{z^2} \erfc(z)$ is the scaled complementary error
function.  Similarly, one gets a simpler expression for $\alpha = -2$:
\begin{equation}
\Upsilon(\mu) = 1 - \frac{\sqrt{\pi} \, \mu \ell_0}{\sqrt{2\nu}} \, \erfcx\bigl(\sqrt{\nu/2} + \mu \ell_0/\sqrt{2\nu}\bigr).
\end{equation}
%
In addition, we find
\begin{equation}
\langle \hat{\ell}\rangle = \int\limits_0^\infty d\ell \, \Psi(\ell) 
= \ell_0 \frac{e^{\nu/|\alpha|} (|\alpha|/\nu)^{1/|\alpha|}}{|\alpha|} \Gamma(1/|\alpha|, \nu/|\alpha|) .
\end{equation}

(ii) For $\alpha = 0$, one has $\Psi(\ell) = (1 + \ell/\ell_0)^{-\nu}$
so that the moment-generating function reads
\begin{equation}  \label{eq:Upsilon_alp0}
\Upsilon(\mu) = \nu e^{\mu \ell_0} (\mu \ell_0)^\nu \Gamma(-\nu, \mu \ell_0).
\end{equation}
In particular, we have
\begin{equation}
\langle \hat{\ell}\rangle = \begin{cases} \displaystyle \frac{\ell_0}{\nu-1} & (\nu > 1), \cr +\infty & (\nu \leq 1). \end{cases}
\end{equation}

(iii) For $\alpha > 0$, we get
\begin{equation}
\Upsilon(\mu) = \frac{\nu}{\alpha} e^{-\nu/\alpha + \mu \ell_0}  \int\limits_0^1 dz \, e^{(\nu/\alpha)z - (\mu \ell_0) z^{-1/\alpha}} .
\end{equation}
In this case, one has $\Upsilon(0^+) = 1 - \Psi(\infty) = 1 -
e^{-\nu/\alpha} < 1$, and $\langle \hat{\ell}\rangle = +\infty$.  A
Taylor expansion of the exponential function $e^{(\nu/\alpha) z}$
yields
\begin{equation}
\Upsilon(\mu) = \nu e^{-\nu/\alpha + \mu \ell_0} \sum\limits_{n=0}^\infty \frac{(\mu \ell_0)^{\alpha(n+1)}}{n!\, (\nu/\alpha)^{-n}}  \,
\Gamma\bigl(-\alpha(n+1), \mu\ell_0\bigr).
\end{equation}
This series is convergent for any $\mu > 0$.

\section{Asymptotic behavior in the encounter-dependent scenario}
\label{sec:Aasympt}

\subsection{Long-time behavior}
\label{sec:Along}

Qualitatively, the long-time behavior of the PDF $\He(t)$ for
encounter-dependent reactivity is related to the small-$p$ behavior of
its Laplace transform $\tilde{\He}(p)$ via Tauberian theorems.  First,
let us rewrite the spectral expansion (\ref{eq:Hp_spectral}) by using
the identity (\ref{eq:identity1}) as
\begin{equation}  \label{eq:Hp_spectral2}
\tilde{\He}(p) = \frac{p}{D|\Omega|} \sum\limits_{k=0}^\infty \frac{\Upsilon(\mu_k^{(p)})}{\mu_k^{(p)}} 
\biggl|\int\limits_{\Omega} V_k^{(p)}\biggr|^2 .
\end{equation}
According to Eqs. (\ref{eq:Steklov_p0}), the leading contribution to
$\tilde{\He}(p)$ as $p\to 0 $ is given by the principal eigenmode, for
which $V_0^{(p)} \approx V_0^{(0)} = 1/\sqrt{|\Gamma|}$, so that
\begin{equation}  \label{eq:Hp_p0}
\tilde{\He}(p) \approx \frac{p}{v} \frac{\Upsilon(\mu_0^{(p)})}{\mu_0^{(p)}} + O(p)  \qquad (p\to 0),
\end{equation}
with $v$ given by Eq. (\ref{eq:v_def}).  We proceed by distinguishing
three situations.

(i) If the moment-generating function $\Upsilon(\mu)$ is nonanalytic
at $0$, one can use the leading-order approximation $\mu_0^{(p)}
\approx p/v$ to get
\begin{equation}  \label{eq:Hp_p0_bis}
\tilde{\He}(p) \approx \Upsilon(p/v)  \qquad (p\to 0),
\end{equation}
that is intended to capture the leading nonanalytic dependence on $p$.

Since $\Upsilon(\mu)$ is the Laplace transform of the threshold
density $\psi(\ell)$, Eq. (\ref{eq:Hp_p0_bis}) suggests that $\He(t)$
and $v\psi(vt)$ share the same leading long-time decay law, at least
at the level of the dominant exponent.  A rigorous justification
generally requires additional Tauberian or large-deviation arguments
(see, e.g., \cite{Forde15} for an analysis of boundary local time in
an interval).  This correspondence does not determine multiplicative
prefactors or subleading corrections.

(ii) If $\Upsilon(\mu)$ is analytic at $0$, the small-$p$ behavior in
Eq.  (\ref{eq:Hp_p0}) preserves the information on the moments but
provides no insight into the long-time behavior.  One needs thus to
search for singularities of $\Upsilon(\mu_0^{(p)})$ in the complex
plane $p\in \C$.  If the rightmost singularity of $\tilde{\He}(p)$ is
a simple pole at $p=p_0<0$, then $\He(t)$ exhibits an exponential
decay proportional to $e^{p_0t}$.  For instance, in the case of
constant reactivity $\kappa_0$, one has $\Upsilon(\mu) = 1/(1+\mu
D/\kappa_0)$, and the pole condition is $\mu_0^{(p_0)} = -\kappa_0/D$.
By the Robin-Steklov duality, this equation determines the principal
eigenvalue of the Robin Laplacian [see Eq. (\ref{eq:Robin_eigen})],
$\lambda_0^{\kappa_0} = -p_0/D$, that controls the long-time behavior:
\begin{equation}  \label{eq:He_Robin_long}
\He(t) \propto e^{-Dt\lambda_0^{\kappa_0}} \qquad (t\to\infty).
\end{equation}
In general, however, $p_0$ can be a multiple pole or even an essential
singularity so that the long-time behavior of $\He(t)$ can include
subleading corrections such as $t^\gamma e^{-Dt\lambda_0^{\kappa_0}}$
or $e^{-Dt\lambda_0^{\kappa_0} + c t^\gamma}$.  These corrections may
be significant, especially at moderately large $t$.  Their rigorous
analysis would require a proper reformulation of the Steklov problem
(\ref{eq:Helm}) for complex-valued $p$ (see \cite{Bogli22}) that is
beyond the scope of this paper.

(iii) Finally, if $\Upsilon(\mu)$ is an entire function of $\mu$, its
singular behavior is only possible at infinity, i.e., when
$|\mu_0^{(p)}|$ diverges.  Such a divergence occurs when $-p/D$
approaches an eigenvalue $\lambda_j^\infty$ of the Dirichlet Laplacian
(see \cite{Grebenkov27}) so that the long-time behavior of $\He(t)$ is
controlled by the smallest eigenvalue $\lambda_0^{\infty}$.
As the eigenfunctions $V_k^{(p)}$ diverge due to the imposed
normalization (\ref{eq:vk_norm}), it is convenient to use the
representation (\ref{eq:Hp_spectral_Mp}).  In the special case of
first-passage times to $\Gamma$, one has the fixed threshold at $\ell
= 0$ so that $\psi(\ell) = \delta(\ell)$ and $\Upsilon(\mu) \equiv 1$.
As a consequence, $p_0 = - D\lambda_0^{\infty}$ is the simple pole of
$\M_p$ (or $\mu_0^{(p)}$), and one retrieves the expected asymptotic
behavior
\begin{equation}  \label{eq:He_Dirichlet_long}
\He(t) \propto e^{-Dt\lambda_0^\infty}  \qquad (t\to\infty).
\end{equation}
These considerations suggest that the principal Dirichlet eigenvalue
determines the leading exponential decay rate when $\Upsilon(\mu)$ is
entire.  However, the precise asymptotic form may depend on the
singularity structure of the operator-valued expression $\M_p
\Upsilon(\M_p)$, which can generate subexponential corrections.

\begin{table*}
\begin{center}
\begin{tabular}{|c|c|c|c|c|c|} \hline
$\mu_*$     & $\Upsilon(\mu)$ & tail of $\psi(\ell)$        & $\kappae(\infty)$ & $\alpha$ & tail of $\He(t)$  \\  \hline
\multirow{2}{*}{$\mu_* = 0$}& nonanalytic  & subexponential    & \multirow{2}{*}{$0$}  & \multirow{2}{*}{$> -1$} 
& stretched-exponential     \\ 
            &  at $\mu = 0$     & or power law     &              &          &  or power law  \\ \hline
\multirow{2}{*}{$0<\mu_* <\infty$} & analytic    & \multirow{2}{*}{exponential} & \multirow{2}{*}{$>0$}  & 
\multirow{2}{*}{$= -1$} & \multirow{2}{*}{exponential}  \\  
                  & near $\mu = 0$ &                  &              &          &              \\ \hline
$\mu_* =\infty$ & entire      & superexponential & $\infty$     & $< -1$   & exponential  \\  \hline
\end{tabular}
\end{center}
\caption{
Summary of the three regimes and the inter-related properties of the
moment-generating function $\Upsilon(\mu)$, the large-$\ell$ behavior
of the PDF $\psi(\ell)$, the large-$\ell$ limit of the
encounter-dependent reactivity $\kappae(\ell)$, and the long-time
behavior of the PDF $\He(t)$.  The exponent of the power-law model
(\ref{eq:kappa_ell}) is also presented.  The case of decreasing
reactivity (first line) can be refined and split into three subcases
according to the value of $\alpha$, see Eq. (\ref{eq:He_power}).}
\label{tab:asympt}
\end{table*}

Since $\Upsilon(\mu)$ is the Laplace transform of the probability
density $\psi(\ell)$, the analyticity of $\Upsilon(\mu)$ can be
directly related to the asymptotic behavior of $\psi(\ell)$.  In order
to classify different regimes, we introduce
\begin{equation}
\mu_* = \sup\biggl\{ \mu > 0 ~:~ \int\limits_0^\infty d\ell \, e^{\mu \ell} \psi(\ell) < \infty \biggr\} 
\end{equation}
(we set $\mu_*=0$ if the integral diverges for every $\mu>0$).  The
value of $\mu_*$ characterizes the exponential-moment properties of
the threshold distribution and provides a useful classification of the
corresponding long-time regimes, as summarized in Table
\ref{tab:asympt}.

To illustrate these distinctions, we consider again the power-law
model (\ref{eq:kappa_ell}), for which the Laplace transform
$\Upsilon(\mu)$ is given by Eq. (\ref{eq:Upsilon_power}). The
preceding arguments motivate the five long-time regimes summarized in
Eq. (\ref{eq:He_power0}).  For decreasing reactivity ($\alpha>-1$),
the approximation $\He(t)\approx v\psi(vt)$ leads to
\begin{equation}  \label{eq:He_power}
\He(t) \simeq \frac{\nu}{t_0} \times \begin{cases} 
e^{-\nu/\alpha}\, (t/t_0)^{-\alpha-1} &  (\alpha > 0), \cr
(t/t_0)^{-\nu-1}  & (\alpha = 0), \cr 
e^{\nu/|\alpha|} (t/t_0)^{|\alpha|-1} & \cr
\quad \times \exp\bigl(-\frac{\nu}{|\alpha|} (t/t_0)^{|\alpha|}\bigr) & (-1< \alpha < 0). \end{cases}
\end{equation}
For $\alpha\geq0$, these expressions provide the leading power-law
asymptotics, whereas for $-1<\alpha<0$, they primarily capture the
leading stretched-exponential behavior on a logarithmic scale, thus
missing subleading corrections.

\subsection{Short-time behavior}
\label{sec:Ashort}

In turn, the short-time asymptotic behavior of $\He(t)$ corresponds to
the large-$p$ behavior of $\tilde{\He}(p)$.  In this limit, many
eigenmodes may provide comparable contributions to the spectral
expansion (\ref{eq:Hp_spectral}), and it is therefore convenient to
use directly Eq. (\ref{eq:Hp_spectral_Mp}).  Since all eigenvalues
$\mu_k^{(p)}$ of $\M_p$ grow as $\sqrt{p}$ (see, e.g.,
\cite{Grebenkov27,Chaigneau24}), the large-$p$ asymptotic behavior is
controlled by the behavior of $\Upsilon(\mu)$ at large $\mu$.  Given
that $\Upsilon(\mu)$ is the Laplace transform of $\psi(\ell)$, one has
to look at the small-$\ell$ behavior of $\psi(\ell)$.

Assuming that the threshold density $\psi(\ell)$ has a finite,
strictly positive limit $\psi(0)$, its Laplace transform satisfies
$\Upsilon(\mu)\sim\psi(0)/\mu$ as $\mu\to\infty$.  Since the smallest
eigenvalue of $\mathcal M_p$ diverges as $p\to\infty$, this asymptotic
relation applies uniformly to its spectrum.  Substituting it into the
operator representation (\ref{eq:Hp_spectral_Mp}), we obtain
\begin{equation}  
\tilde{\He}(p) \approx \frac{D}{p|\Omega|}\, |\Gamma| \psi(0) \qquad (p\to \infty),
\end{equation}
and thus
\begin{equation}
\He(t) \approx v\, \psi(0) \qquad (t\to 0).
\end{equation}
For the power-law model (\ref{eq:kappa_ell}), one has $\psi(0) =
\nu/\ell_0$ so that $\He(t) \approx \kappa_0 |\Gamma|/|\Omega|$.  This
result is consistent with the initial reactive flux from a uniform
concentration: at short times, the accumulated boundary local time is
small, and the encounter-dependent reactivity is effectively equal to
its initial value $\kappae(0) = \kappa_0$.

\section{Adiabatic approximation}
\label{sec:Aadiabatic}

We justify the adiabatic approximation of Sec. \ref{sec:adiabatic} by
estimating the contribution of nonprincipal instantaneous eigenmodes.
Denoting the coupling coefficients in Eq. (\ref{eq:cn_eqs}) by $A_{nj}
= (U_n^{\kappa(t)},\partial_\kappa U_j^{\kappa(t)})_{L^2(\Omega)}$,
differentiation of the orthonormality relations, $(U_n^{\kappa},
U_j^{\kappa})_{L^2(\Omega)} = \delta_{nj}$, gives
\begin{equation}  \label{eq:A-antisym}
A_{nj} = - A_{jn}, \qquad A_{nn} = 0.
\end{equation}
Thus, the variation of the instantaneous eigenbasis produces no
diagonal coupling.   In addition, taking the derivative of
Eqs. (\ref{eq:Robin_eigen}) with respect to $\kappa$, projecting on
$U_n^{\kappa(t)}$, applying Green's identity and using the boundary
conditions, one gets
\begin{equation*}
\bigl(\lambda_n^{\kappa(t)} - \lambda_j^{\kappa(t)}\bigr) A_{nj} = \delta_{jn} \partial_\kappa \lambda_j^{\kappa(t)} 
-\frac{1}{D} \bigl(U_n^{\kappa(t)}, U_j^{\kappa(t)}\bigr)_{L^2(\Gamma)} \,,
\end{equation*}
yielding an alternative representation for $A_{nj}$.

In a first step, we factor out the instantaneous decay of the
principal mode.  Setting
\begin{equation}  \label{eq:def-bn}
c_n(t) = b_n(t) \exp\left[ -D\int\limits_{t_*}^{t} dt'\,\lambda_0^{\kappa(t')} \right]
\end{equation}
for some large enough timescale $t_*$, we get
\begin{equation} \label{eq:b-system-general}
\dot b_m =  -\Delta_m(t) b_m  - \dot\kappa\sum_{n=0}^\infty A_{mn}\, b_n,
\end{equation}
where
\begin{equation}
\Delta_m(t) =  D\bigl[\lambda_m^{\kappa(t)} - \lambda_0^{\kappa(t)}\bigr], \qquad  \Delta_0=0.
\end{equation}

For a bounded, connected domain with sufficiently regular boundary,
the principal eigenvalue of the Robin Laplacian is simple and remains
separated from the rest of the spectrum.  Since the eigenvalues depend
continuously on $\kappa\in[0,\infty]$, including the Neumann and mixed
Dirichlet-Neumann limits, there exists a constant $g_*>0$ such that
\begin{equation}  \label{eq:uniform-gap}
\Delta_m(t)\geq g_*>0,  \qquad m\geq1,\quad t\geq t_*.
\end{equation}

Let
\begin{equation}
R^2(t)=\sum\limits_{m=1}^\infty |b_m(t)|^2
\end{equation}
measure the total amplitude of the nonprincipal modes.  Multiplying
Eq. \eqref{eq:b-system-general} by $b_m$, summing over $m$ and using
the antisymmetry relation \eqref{eq:A-antisym}, one obtains
\begin{align}
\frac{1}{2}\frac{dR^2}{dt} &=  -\sum\limits_{m=1}^\infty \Delta_m|b_m|^2 - \dot\kappa\,b_0 \sum\limits_{m=1}^\infty A_{m0} b_m 
\end{align}
(the terms coupling the excited modes among themselves cancel because
$A_{mn}$ is antisymmetric).  Using Eq. (\ref{eq:uniform-gap}) and the
Cauchy-Schwarz inequality
\begin{equation*}
\left|\sum\limits_{m=1}^\infty A_{m0} b_m \right| \leq R(t) \biggl( \sum\limits_{m=1}^\infty |A_{m0}|^2 \biggr)^{1/2} ,
\end{equation*}
we obtain
\begin{equation} \label{eq:R-bound}
\dot R \leq -g_* R + q(t)|b_0|,
\end{equation}
where
\begin{equation}
q(t) = |\dot\kappa(t)| \biggl( \sum\limits_{m=1}^\infty |A_{m0}|^2 \biggr)^{1/2}.
\end{equation}
Since $A_{00}=0$, completeness of the instantaneous eigenbasis yields
\begin{equation}   \label{eq:q-def}
q(t) = |\dot\kappa(t)|\, \|\partial_\kappa U_0^{\kappa(t)}\|_{L^2(\Omega)} .
\end{equation}
Integration of \eqref{eq:R-bound} gives
\begin{equation}   \label{eq:R-convolution}
R(t) \leq R(t_*)e^{-g_*(t-t_*)}  +  \int\limits_{t_*}^{t} dt' \, e^{-g_*(t-t')} q(t')|b_0(t')| .
\end{equation}
The first term represents an arbitrary initial contribution of the
higher modes and decays exponentially on the spectral-gap timescale
$g_*^{-1}$.  The second term describes how the higher modes are
continuously generated by the variation of the instantaneous
eigenbasis.

It remains to verify that the principal amplitude $b_0$ varies slowly.
For any finite $\kappa$ and a bounded domain with sufficiently smooth
boundary, the principal eigenfunction $U_0^{\kappa}$ does not change
sign in $\Omega \cup \pa$ and thus can be chosen positive.  Actually
the Hopf lemma implies that $U_0^{\kappa} > 0$ \cite{Levitin}.  In
addition, positivity of $C(\x,t)$ ensures
\begin{equation}
c_0(t) = \bigl(C(\cdot,t), U_0^{\kappa(t)}\bigr)_{L^2(\Omega)} > 0 \quad \Rightarrow \quad b_0(t) > 0,
\end{equation}
where we set $C_0 = 1$ for brevity.  Differentiating $c_0(t)$ by
$\kappa$ and using self-adjointness of the Robin Laplacian at fixed
$t$, we have
\begin{equation}\label{eq:proj}
\dot c_0=- D\lambda_0^{\kappa(t)} c_0 + \dot\kappa(t) \bigl(C(\cdot,t), \partial_\kappa U_0^{\kappa(t)}\bigr)_{L^2(\Omega)}.
\end{equation}
Consequently,
\begin{equation}\label{eq:logder}
\partial_t \ln b_0(t) = \dot\kappa(t)
\frac{\bigl( C(\cdot,t), \partial_\kappa U_0^{\kappa(t)}\bigr)_{L^2(\Omega)}}{\bigl(C(\cdot,t), U_0^{\kappa(t)}\bigr)_{L^2(\Omega)}}.
\end{equation}
Assume that
\begin{equation}  \label{eq:M_def}
M(\kappa)=\left\|\frac{\partial_\kappa U_0^\kappa}{U_0^\kappa}\right\|_{L^\infty(\Omega)}
\end{equation}
is finite for any $\kappa$.  Since $|\partial_\kappa
U_0^{\kappa}/U_0^{\kappa}| \leq M(\kappa)$ pointwise and $U_0^{\kappa}$
is positive, multiplication by $C(\x,t) U_0^{\kappa}$ of both sides of
this inequality and integration over $\Omega$ yield
\begin{equation*}
\left|\bigl( C(\cdot,t), \partial_\kappa U_0^{\kappa}\bigr)_{L^2(\Omega)}\right| 
\leq M(\kappa) \bigl(C(\cdot,t), U_0^{\kappa}\bigr)_{L^2(\Omega)}, 
\end{equation*}
from which
\begin{equation}\label{eq:basic}
\left|\partial_t \ln b_0(t)\right|\leq\eta(t)= |\dot\kappa(t)| M(\kappa(t)).
\end{equation}
Thus, for $t_*\leq t'\leq t$,
\begin{equation}\label{eq:ratio}
\frac{b_0(t')}{b_0(t)}\leq \exp \left(\int_{t'}^t dt'' \, \eta(t'')\right).
\end{equation}
If 
\begin{equation}  \label{eq:omega_def}
\omega = \int\nolimits_{t_*}^\infty dt' \, \eta(t') < \infty, 
\end{equation}
then $b_0(t') \leq e^\omega b_0(t)$.  Substituting this inequality
into Eq. (\ref{eq:R-convolution}), we get
\begin{equation}  \label{eq:R-convolution2}
\frac{R(t)}{b_0(t)} \leq e^\omega \biggl[\frac{R(t_*)}{b_0(t_*)} e^{-g_*(t-t_*)} + \int\limits_{t_*}^t dt' \, e^{-g*(t-t')} q(t')\biggr].
\end{equation}
If $q(t)$ varies slowly on the relaxation timescale $g_*^{-1}$, as is
the case for the power-law reactivities considered below, the
convolution in Eq. \eqref{eq:R-convolution2} is asymptotically
controlled by $q(t)/g_*$.  Consequently,
\begin{equation}   \label{eq:R-adiabatic}
\frac{R(t)}{b_0(t)} = O\bigl(\epsilon(t) \bigr) +  O(e^{-g_*t}),
\end{equation}
where we have identified the natural adiabatic parameter
\begin{equation}  \label{eq:adiabatic-parameter}
\epsilon(t) = \frac{|\dot\kappa(t)| \,  \|\partial_\kappa U_0^{\kappa(t)}\|_{L^2(\Omega)}}
{D[\lambda_1^{\kappa(t)} - \lambda_0^{\kappa(t)}] } .
\end{equation}
It has a simple interpretation: $\epsilon(t)$ measures the change of
the instantaneous principal eigenfunction during one relaxation time
associated with the spectral gap.  Thus, under the assumptions leading
to Eq. (\ref{eq:R-adiabatic}), if $\epsilon(t)\to0 $ as $t\to\infty$,
the nonprincipal component becomes asymptotically small compared with
the principal mode $b_0(t)$.  

This completes our justification of the adiabatic approximation, up to
two technical assumptions that $M(\kappa)$ and $\omega$ in
Eqs. (\ref{eq:M_def}, \ref{eq:omega_def}) are finite.  
For reactivities exhibiting a power-law decrease,
Eq. (\ref{eq:omega_def}) follows if $M(\kappa)$ is bounded near
$\kappa = 0$, since $\int_{t_*}^\infty |\dot\kappa(t)|\,dt<\infty$.

When $\kappa(t)\to\infty$, the control of the principal amplitude
through Eqs. (\ref{eq:M_def} -- \ref{eq:omega_def}) becomes more
delicate because the principal eigenfunction vanishes on the reactive
boundary in the Dirichlet limit.  Fortunately, the leading exponential
decay rate can be established independently, without these
assumptions, by using the comparison principle.  For any $K > 0$,
there is $T$ large enough such that $\kappa(t) \geq K$ for any $t >
T$.  The comparison principle then bounds the subsequent survival by
that of the constant-$K$ Robin problem, up to a positive
multiplicative factor.  Conversely, the absorbing Dirichlet problem
provides a lower bound on survival:
\begin{equation}  \label{eq:lambda0_dominance}
D\lambda_0^K + o(1) \leq - \frac{\ln S(t)}{t} \leq D \lambda_0^\infty + o(1)\qquad (t > T).
\end{equation} 
Taking first the long-time limit and then letting $K\to\infty$, so
that $\lambda_0^K \to \lambda_0^\infty$, one thus establishes the
asymptotic relation (\ref{eq:St_Dirichlet}).  Remarkably, this
argument does not rely on the adiabatic approximation and requires
neither a slow variation, nor monotonicity of $\kappa(t)$; it only
assumes that $\kappa(t)\to\infty$ as $t\to\infty$.

We now estimate the adiabatic parameter $\epsilon(t)$ for the
power-law reactivity model (\ref{eq:kappa_model}).  If $\kappa(t)\to
0$, regular perturbation theory gives $\|\partial_\kappa
U_0^{\kappa}\|_{L^2(\Omega)} =O(1)$, while the spectral gap tends to
the first nonzero Neumann eigenvalue $\lambda_1^{0}$.  Thus, for
$\kappa(t)\propto t^{-\gamma}$ with $\gamma>0$, one has
\begin{equation}
\epsilon(t) = O(t^{-\gamma-1}) \to 0.
\end{equation}
Conversely, for $\kappa(t)\to\infty$,
\begin{equation}
U_0^{\kappa}  = U_0^{\infty} + O(\kappa^{-1}), \qquad  \partial_\kappa U_0^{\kappa} = O(\kappa^{-2}),
\end{equation}
whereas the spectral gap tends to the finite Dirichlet-Neumann gap.
For $\kappa(t)\propto t^\gamma$ with $\gamma>0$, this gives
\begin{equation}
\epsilon(t) = O\left(\frac{|\dot\kappa|}{\kappa^2}\right) = O(t^{-\gamma-1}) \to 0.
\end{equation}
For increasing reactivity, this estimate establishes $\epsilon(t) \to
0$. The logarithmic survival asymptotic follows directly from
Eq. (\ref{eq:lambda0_dominance}).

\begin{figure}
\centering
\includegraphics[width=0.49\columnwidth]{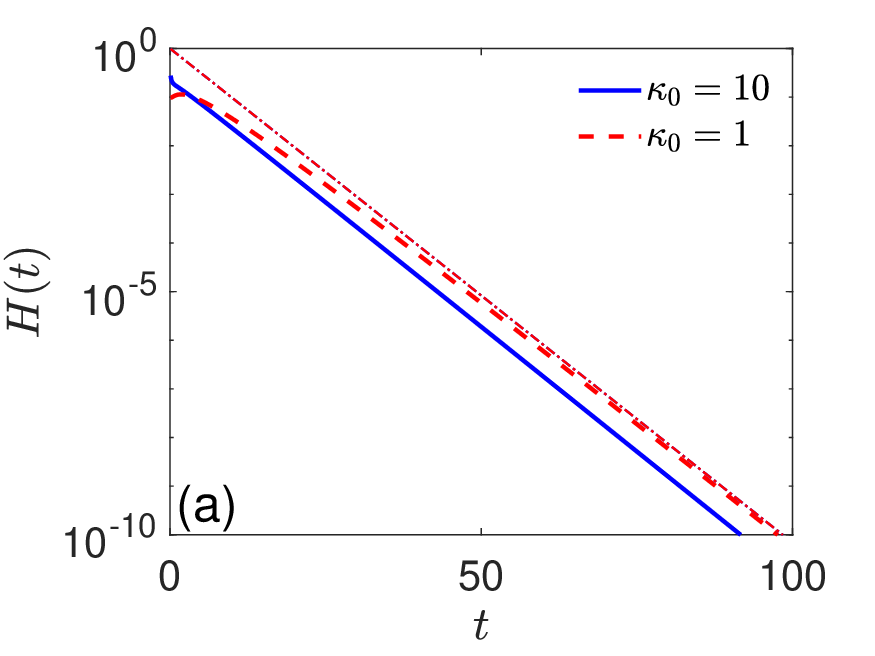} 
\includegraphics[width=0.49\columnwidth]{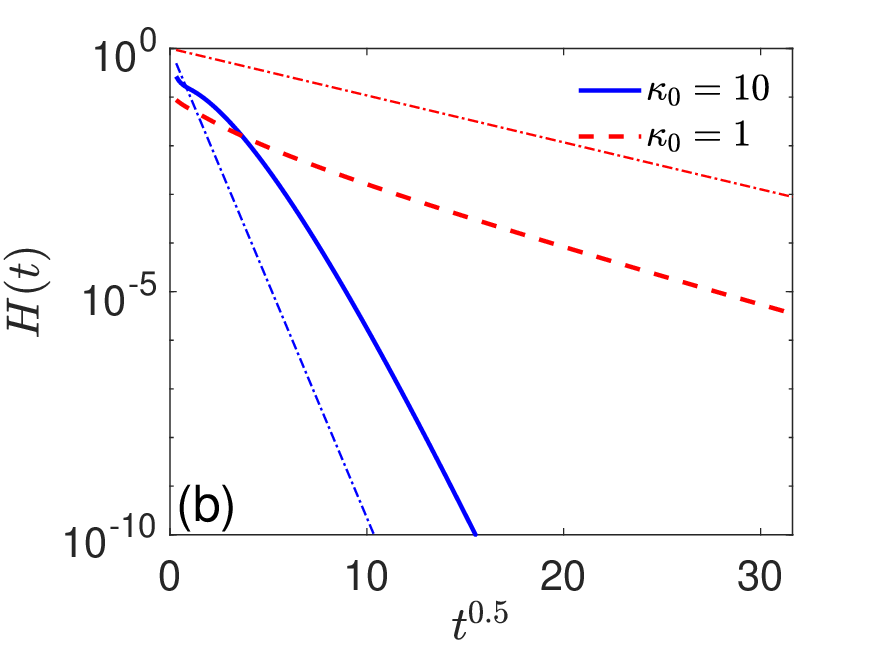} 
\includegraphics[width=0.49\columnwidth]{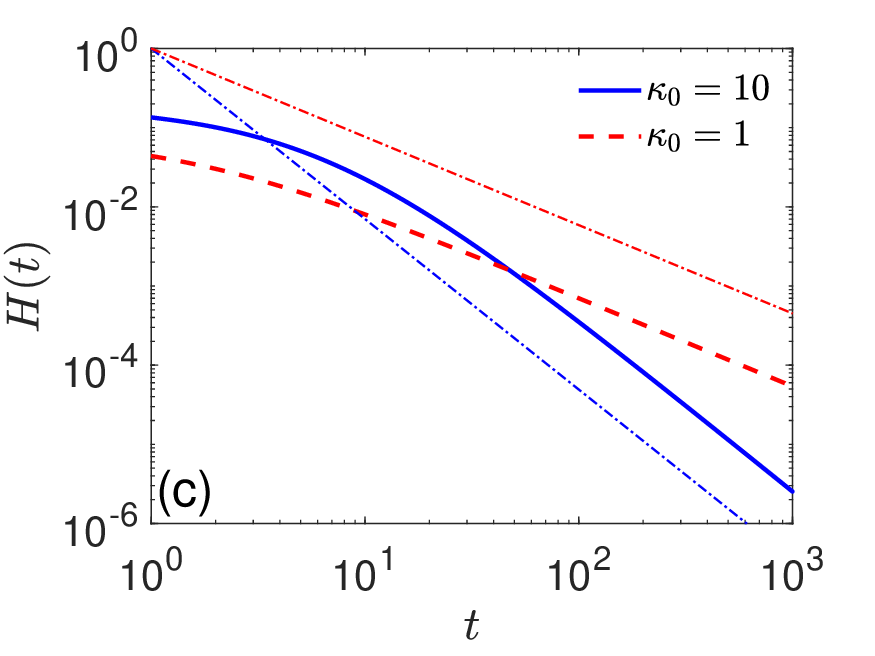} 
\includegraphics[width=0.49\columnwidth]{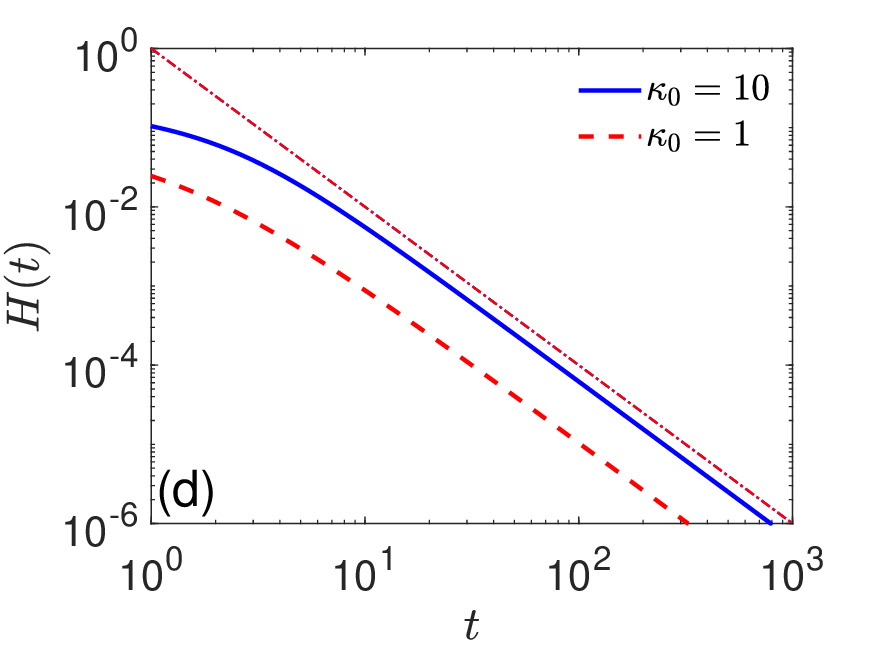} 
\caption{
Long-time asymptotic behavior of the volume-averaged PDF $H(t)$ of the
first-reaction time on a spherical target of radius $R = 1$, with
time-dependent reactivity $\kappa(t)$ from Eq. (\ref{eq:kappa_model}),
and surrounded by a reflecting concentric sphere of radius $L = 3$.
Solid and dashed lines present $H(t)$ obtained by a finite-difference
method (Appendix \ref{sec:FDM}) for $\kappa_0 = 10$ and $\kappa_0 =
1$, respectively.  Thin dash-dotted lines show the asymptotic
relations in Eq. (\ref{eq:Ht_power}), with unit multiplicative
amplitudes (such as unknown prefactor $A_\alpha$).  We set $D = 1$,
$t_0 = 1$, and four values of $\alpha$: $-2$ (panel (a)), $-0.5$
(panel (b)), $0$ (panel (c)), and $1$ (panel (d)).}
\label{fig:Ht_asympt}
\end{figure}

Figure \ref{fig:Ht_asympt} illustrates the predicted asymptotic
behavior (\ref{eq:Ht_power}) for the power-law reactivity model
(\ref{eq:kappa_model}) and shell geometry.  When $\alpha = -2$ (panel
(a)), the growing reactivity $\kappa(t)$ results in an exponential
decay of $H(t)$ controlled by the principal Dirichlet(-Neumann)
Laplacian eigenvalue (here, $\lambda_0^\infty \approx 0.2340$),
independently of the initial reactivity $\kappa_0$.  Indeed, one sees
that both curves (shown with the logarithmic scale of the vertical
axis) approach the slope $-D\lambda_0^\infty$.  The initial reactivity
$\kappa_0$ affects the subleading behavior.  In particular, the
$O(\kappa^{-1})$ correction to the instantaneous eigenvalue can
generate an $O(\ln t)$ contribution to the exponent for the linearly
increasing reactivity used here.  When $\alpha = -0.5$ (panel (b)),
the stretched-exponential decay of $H(t)$ is visualized by plotting
$H(t)$ as a function of $t^{0.5}$, with the logarithmic scale of the
vertical axis.  The initial reactivity $\kappa_0$ determines $\nu$ and
thus the slope of the shown asymptotic straight lines.  When $\alpha =
0$, the PDF $H(t)$ exhibits a power-law decay, with the exponent
controlled by $\kappa_0$, as clearly seen on panel (c).  Finally, when
$\alpha = 1$, the power-law decay of $H(t)$ is controlled by the
exponent $\alpha$, whereas the initial reactivity $\kappa_0$ affects
only the prefactor.  These four examples illustrate the validity of
our asymptotic predictions (\ref{eq:Ht_power}) but also reveal the
importance of finding yet unknown subleading terms to get an accurate
description.

\section{Numerical methods}
\label{sec:Anumerics}

In this Appendix, we briefly describe the numerical methods employed
in the manuscript and their validation.

\subsection{Time-dependent reactivity}
\label{sec:direct}

To get the PDF $H(t)$ of the FRT on a reactive surface $\Gamma$ with a
given time-dependent reactivity $\kappa(t)$, we employ Monte Carlo
simulations.  For a given confining domain with a smooth boundary, one
first generates a uniformly distributed starting point $\x_0$ and then
runs the standard walk-on-spheres (WOS) algorithm \cite{Muller56} for
sampling a sequence of points $\{\x_0,\x_1,\x_2,\ldots\}$ on a random
trajectory in the bulk.  Iteratively, from the current point $\x_j$,
one determines the distance $\rho_j = |\x_j - \pa|$ to the boundary
$\pa$ and generates the new point $\x_{j+1}$ uniformly on the sphere
of radius $\rho_j$ centered at $\x_j$.  The time counter is
incremented as $t_{j+1} = t_j + (\rho_j^2/D) \tau_c$, where $\tau_c$
is the random escape time from the center of the unit ball, which has
a well-known distribution (see, e.g., \cite{Ye25}).  As the jump
distance $\rho_j$ is reduced as the particle approaches the boundary,
such a process would never hit the boundary.  In order to account for
reflections and eventual reactions, we introduce a thin boundary layer
of width $\ve$.  The above simulation is performed until the distance
$\rho_k$ becomes smaller than $\ve$; at this moment, the particle is
considered to be inside that layer.

In standard Monte Carlo techniques, a refined simulation of the random
trajectory inside the boundary layer is continued, until the particle
either leaves the layer, or reacts on $\Gamma$.  Such simulations are
generally very long and time-consuming because random displacements
are usually ensured to be much smaller than $\ve$.  Here, we adapt the
escape-from-a-layer approach \cite{Ye25}, in which a long simulation
of the trajectory inside a thin layer is replaced by a single escape
or reaction event.  To reduce the computational cost, we approximate
the random duration of each boundary-layer event by its conditional
mean, as given below.  This approximation becomes increasingly
accurate as the layer width $\ve$ decreases, and its effect on the
numerical results has been assessed by varying $\ve$.  As the smooth
boundary is locally flat, the probability of escaping this layer
without hitting the boundary is $\rho_k/\ve$.  Generating a uniformly
distributed random variable $\eta$, one can therefore choose between
two outcomes:

(i) if $\eta < \rho_k/\ve$, the particle exits the boundary layer
without hitting the boundary $\pa$; in this case, the particle is
relocated on the inner border of the boundary layer, $\x_{k+1} =
\x_k - (\ve - \rho_k)\n(\x_k)$, where $\n(\x_k)$ is the unit normal
vector at the boundary point, which is the closest to $\x_k$, and the
time counter is increased as $t_{k+1} = t_k + (\ve^2 -
\rho_k^2)/(6D)$, where the increment is the conditional mean escape
time from that layer (see details in \cite{Ye25}); the WOS algorithm
is then resumed.

(ii) if $\eta > \rho_k/\ve$, the particle hits the boundary and
attempts to react on it with the probability \cite{Grebenkov03}
\begin{equation}  \label{eq:pt_exact}
p_{t_k} = 1 - \frac{1}{1 + \ve \kappa(t_k)/D} \, .  
\end{equation}
For this purpose, one generates another uniform random variable
$\eta'$: if $\eta' < p_{t_k}$ and the closest boundary point belongs
to the reactive surface $\Gamma$, the reaction attempt is successful,
and the simulation is stopped; otherwise, the reaction attempt fails,
the particle is relocated on the inner border of the boundary layer as
earlier, the time counter is increased as $t_{k+1} = t_k + (\ve +
\rho_k)(3\ve - \rho_k)/(6D)$, and the WOS algorithm is resumed.

Once the simulation of a trajectory is stopped, the reaction time
$t_k$ is recorded.  One then repeats this simulation independently $N$
times to get an empirical statistics of first-reaction times.  In the
simulations presented here, we used $N = 10^6$ and $\ve = 10^{-3}$.

The above simulation scheme assumes that the reactivity $\kappa(t)$
remains constant during each exploration of the thin boundary layer,
i.e., variations of $\kappa(t)$ during the diffusion timescale
$\ve^2/D$ are negligible; in other words, we assume that
$|\dot{\kappa}(t)|/\kappa(t)\ll D/\ve^2$.  This condition is satisfied
for all considered models.  Note also that if the layer width $\ve$ is
very small, the reaction probability $p_{t_k}$ could be linearized as
$\ve \kappa(t_k)/D$, thus recovering the probability of reaction
$p_{\rm e}$ discussed at the beginning of Sec. \ref{sec:encounter}.
However, the use of Eq. (\ref{eq:pt_exact}) is preferred to deal
numerically with increasing reactivities.

When the reactivity $\kappa(t)$ decays as $t^{-1-\alpha}$ with an
exponent $\alpha > 0$, a positive fraction of particles may never
react (see Sec. \ref{sec:time_proba}).  To avoid simulating
arbitrarily long trajectories, we introduce a maximal simulation time
$T_{\rm max}$ (we set $T_{\rm max} = 10^4$).  Trajectories that have
not reacted by this time are treated as right-censored, i.e., their
reaction times are known only to exceed $T_{\rm max}$.  Such censoring
does not affect the estimated PDF of the first-reaction time for $t<
T_{\rm max}$, but it prevents a direct estimation of the no reaction
probability.

\subsection{Encounter-dependent simulations}
\label{sec:MC_encounter}

In the encounter-dependent scenario, we can access the PDF $\He(t)$
analytically, at least for the shell domain, as described below in
Appendix \ref{sec:shell}.  However, we perform Monte Carlo simulations
even in this case for an independent validation of analytical results
and to access the accuracy of the involved numerical computations
(such as inverse Laplace transforms).

An efficient WOS algorithm for simulating surface reactions with a
given encounter-dependent reactivity $\kappae(\ell)$ was proposed in
\cite{Ye25} (see also Appendix \ref{sec:direct}).  The specific
feature of this algorithm consists in generating the boundary local
time accumulated inside a thin boundary layer at each escape event as
$\ell_{t_{k+1}} = \ell_{t_k} + d\ell_{t_k}$, where the increment
$d\ell_{t_k}$ is generated from the exponential distribution with the
mean $\ve$ (conditional on hitting the boundary before exiting the
layer).  The algorithm is stopped when $\ell_{t_k}$ exceeds a random
threshold $\hat{\ell}$.

For the power-law model (\ref{eq:kappa_ell}), the threshold can be
easily generated by inverting the relation $\eta = \Psi(\ell)$, with
$\Psi(\ell)$ given by Eq. (\ref{eq:Psi_model}):
\begin{equation}  \label{eq:hatell_eta}
\hat{\ell} = \begin{cases} \displaystyle \ell_0\biggl( \biggl(1 + \frac{\alpha}{\nu}\ln \eta\biggr)^{-1/\alpha} - 1\biggr)
& (\alpha \ne 0), \cr   \ell_0 (\eta^{-1/\nu} - 1)  & (\alpha = 0), \end{cases}
\end{equation}
where $\eta$ is the uniform random variable on $(0,1)$.  For
$\alpha>0$, Eq. (\ref{eq:hatell_eta}) applies when
$\eta>e^{-\nu/\alpha}$, whereas $\hat\ell=\infty$ is assigned when
$\eta<e^{-\nu/\alpha}$.  The latter event occurs with probability
$e^{-\nu/\alpha}$ and corresponds to a particle that never reacts.
Such trajectories can be identified without simulating their
diffusion, allowing the no reaction probability to be determined
directly, without introducing a maximal simulation time.

\subsection{Alternative Monte Carlo simulations} 
\label{sec:MC_proba}

We also implemented an alternative Monte Carlo simulation of the
time-dependent reactivity model, based on the probabilistic
interpretation introduced in Sec. \ref{sec:time_proba}.  Specifically,
we construct the reaction functional $L_t$ through the discrete
updates $L_{t_{k+1}} = L_{t_k}+\kappa(t_k)d\ell_{t_k}/D$, where
$d\ell_{t_k}$ is the boundary local-time increment generated by the
escape-from-a-layer algorithm \cite{Ye25} (see also Appendix
\ref{sec:MC_encounter}).  This update approximates the stochastic
integral in Eq. (\ref{eq:Lt_def}) by treating $\kappa(t)$ as constant
during each boundary-layer event.  The trajectory is stopped according
to Eq. (\ref{eq:tau_defL}) when $L_{t_k}$ exceeds an independent
exponential threshold $\hat{L}$ with mean $1$.  Repeating this
procedure for $N$ independent trajectories yields an empirical
estimate of the PDF $H(t)$ of FRT.  As in Appendix
\ref{sec:direct}, trajectories that have not reacted by the maximal
simulation time $T_{\rm max}$ are treated as right-censored.

\subsection{Encounter-based computation for spherical shell}
\label{sec:shell}

Our illustrations of theoretical results are provided for a spherical
reactive surface $\Gamma$ of radius $R$, surrounded by a concentric
impermeable sphere $\pa_0$ of radius $L$.  For such
rotationally-invariant domains, the separation of variables implies
that $V_k^{(p)}$ on $\Gamma$ are spherical harmonics, which are
orthogonal to the constant function $V_0^{(p)} = 1/\sqrt{|\Gamma|}$.
As a consequence, only the spherically symmetric eigenmode with $k =
0$ remains in Eq. (\ref{eq:Up}), yielding
\begin{equation}  \label{eq:Ut_sphere}
\tilde{U}(\ell,p) = \frac{D |\Gamma|}{|\Omega| p} \, \mu_0^{(p)} e^{-\mu_0^{(p)}\ell} \,,
\end{equation}
from which
\begin{equation}  \label{eq:Ht_sphere}
\tilde{\He}(p) = \frac{D |\Gamma|}{|\Omega| p} \, \mu_0^{(p)} \Upsilon(\mu_0^{(p)}) \, ,
\end{equation}
where the smallest eigenvalue is \cite{Grebenkov20b}
\begin{equation} \label{eq:mu0_shell}
\mu_0^{(p)} = \frac{(q L-1)(q R+1) - e^{-2q(L-R)}(q L+1)(q R-1)}{R[q L-1 + e^{-2q(L-R)}(q L+1)]} \,,
\end{equation}
with $q = \sqrt{p/D}$.

In order to get the PDF $\He(t)$ in time domain, one needs to perform
the inverse Laplace transform numerically.  For this purpose, we use
the Talbot algorithm \cite{Talbot79}.  There are at least two possible
implementations.  The first one consists in evaluating the inverse
Laplace transform of Eq. (\ref{eq:Ht_sphere}) with $\Upsilon(\mu)$ for
a given model.  However, in the considered case of the power-law
model, the moment-generating function $\Upsilon(\mu)$ is also given by
the integral (\ref{eq:Upsilon_power}) that makes the application of
the Talbot algorithm less efficient.  We therefore use a different
implementation.

For a sufficiently large $\ell_{\rm max}$ and $K$, we introduce a
nonuniform mesh $l_k = \ell_s [(1 + kh)^2 - 1]$, with $k =
0,1,\ldots,K$, where $h = (\sqrt{1+\ell_{\rm max}/\ell_s}-1)/K$ and
$\ell_s$ is a characteristic length scale controlling the
concentration of mesh points near $\ell = 0$ (its choice affects the
numerical resolution but not the converged result; in practice, one
may set $\ell_s$ to $\ell_0$ or $R$).  In this way, the interval
$(0,\ell_{\rm max})$ is discretized such that $l_0 = 0$, $l_K =
\ell_{\rm max}$, and the mesh points are denser at smaller $\ell$.  We
apply the trapezoidal rule to approximate the integral in
Eq. (\ref{eq:Ht_Ut}):
\begin{equation}
\He(t) \approx \sum\limits_{k=0}^K w_k\, \psi(l_k)\, U(l_k,t),
\end{equation}
with the standard trapezoidal weights
\begin{equation}
w_0 = \frac{l_1-l_0}{2}, \quad w_k = \frac{l_{k+1}-l_{k-1}}{2}, \quad
w_K = \frac{l_K - l_{K-1}}{2} .
\end{equation}
The PDF $\He(t)$ can thus be sampled at the desired set of times
$t_1,\ldots,t_M$ via a single matrix multiplication, once the matrix
elements $U(l_k,t_m)$ are found by inverting Eq. (\ref{eq:Ut_sphere})
via the Talbot algorithm.  Since $U(\ell,t)$ depends only on the
confining domain, this computation has to be performed only once, and
the stored elements $U(l_k,t_m)$ can be loaded for each computation of
$\He(t)$ for the desired $\psi(\ell)$.  At very long times ($t
\gtrsim 600$), the Gaussian approximation (\ref{eq:U_Gauss}) was
used.

In the examples considered, we set $K=1000$ and $\ell_{\rm max} = 200$
(with lengths expressed in units of $R$).  We checked numerical
convergence by refining the local-time mesh.  The results were also
compared with independent Monte Carlo simulations (Appendix
\ref{sec:MC_encounter}).  An additional consistency check was provided
by the exact identity
\begin{equation}
\frac{D|\Gamma|}{|\Omega|} = \partial_t \langle \ell_t \rangle = \int\limits_0^\infty d\ell \, U(\ell,t)
\approx \sum\limits_{k=0}^K w_k \, U(l_k,t),
\end{equation}
which was verified at the considered times.

\begin{figure}
\centering
\includegraphics[width=\columnwidth]{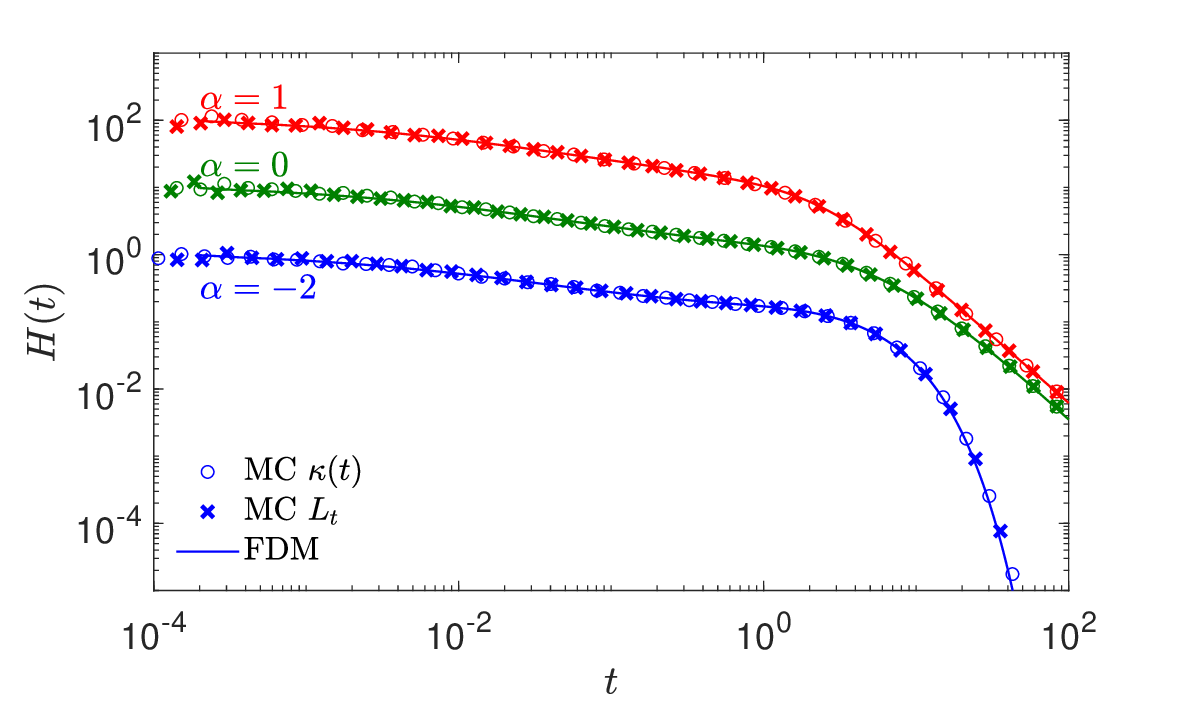} 
\includegraphics[width=\columnwidth]{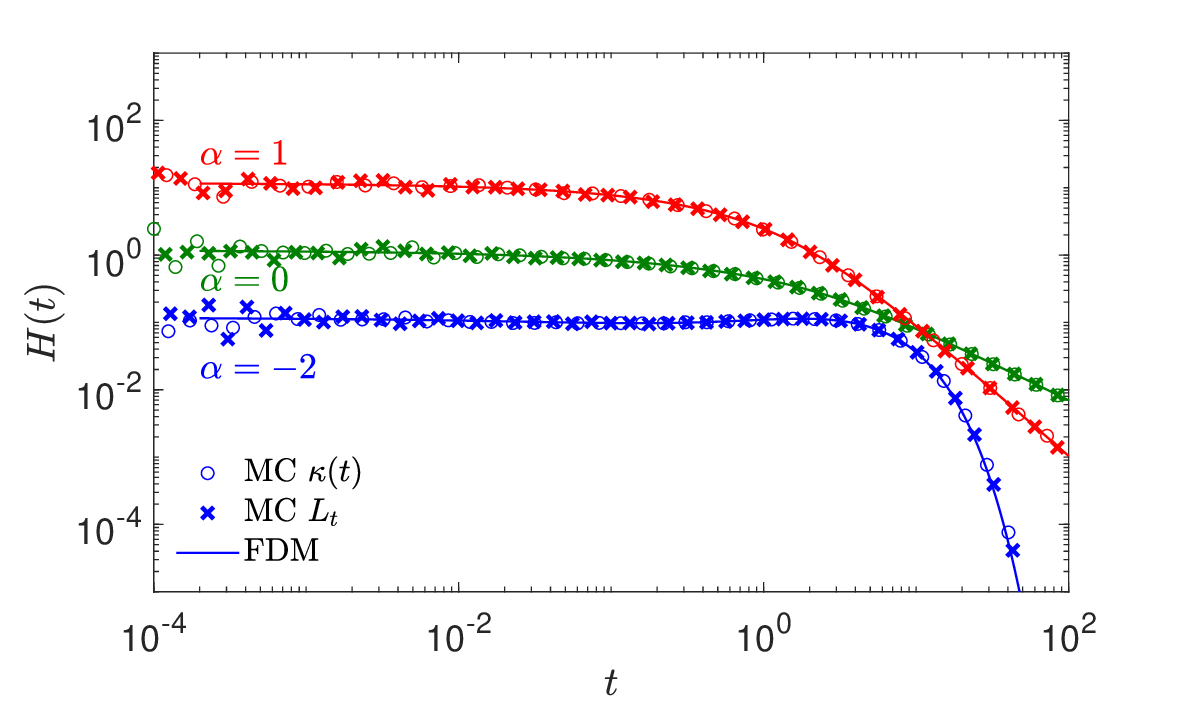} 
\caption{
Comparison of three numerical methods for computing volume-averaged
PDF $H(t)$ of the first-reaction time on a spherical target of radius
$R = 1$, with time-dependent reactivity $\kappa(t)$ from
Eq. (\ref{eq:kappa_model}), and surrounded by a reflecting concentric
sphere of radius $L = 3$.  Circles show the empirical histogram from
direct Monte Carlo simulations (Appendix \ref{sec:direct}); crosses
present the empirical histogram from Monte Carlo simulations with the
exponential stopping condition on $L_t$ (Appendix \ref{sec:MC_proba});
solid lines illustrate the results of finite-difference method
(Appendix \ref{sec:FDM}).  We set $D = 1$, $t_0 = 1$, two values
$\kappa_0 = 10$ (top) and $\kappa_0 = 1$ (bottom), and three values of
$\alpha$: $-2$, $0$, and $1$.  For better visibility, the PDFs for
$\alpha = 0$ (in green) and $\alpha = 1$ (in red) are shifted along
the vertical axis by multiplication by factors $10$ and $100$,
respectively.}
\label{fig:validation}
\end{figure}

\begin{figure}
\centering
\includegraphics[width=\columnwidth]{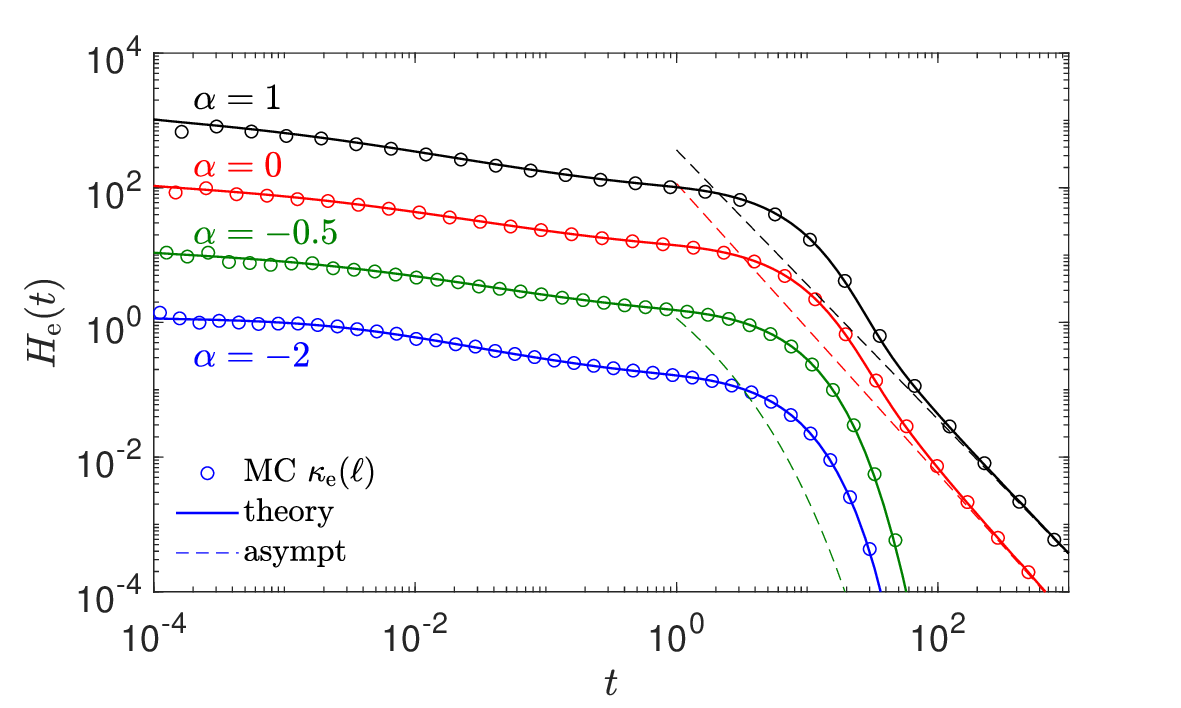} 
\includegraphics[width=\columnwidth]{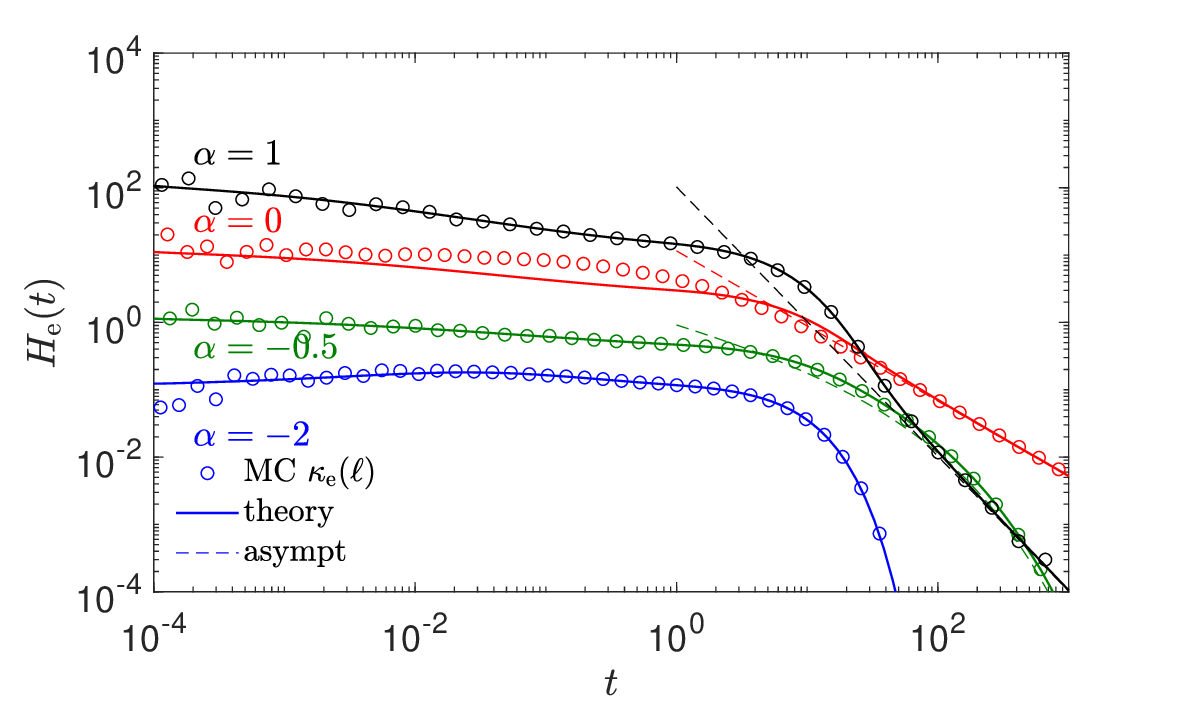} 
\caption{
Comparison of theoretical predictions and numerical results for
volume-averaged PDF $\He(t)$ of the first-reaction time on a spherical
target of radius $R = 1$, with encounter-dependent reactivity
$\kappae(\ell)$ from Eq. (\ref{eq:kappa_ell}) and surrounded by a
reflecting concentric sphere of radius $L = 3$.  Circles show the
empirical histogram from encounter-based Monte Carlo simulations
(Appendix \ref{sec:MC_encounter}), whereas solid lines illustrate the
theoretical prediction (\ref{eq:Ht_Ut}) (see Appendix
\ref{sec:shell}).  Dashed lines indicate the asymptotic behavior
(\ref{eq:He_power}) for $\alpha > -1$.  We set $D = 1$, $t_0 = 1$ (and
thus $\ell_0 = |\Gamma| Dt_0/|\Omega|$), two values $\kappa_0 = 10$
(top) and $\kappa_0 = 1$ (bottom), and four values of $\alpha$: $-2$,
$-0.5$, $0$, and $1$.  For better visibility, the PDFs for $\alpha =
-0.5$ (in green), $\alpha = 0$ (in red), and $\alpha = 1$ (in black)
are shifted along the vertical axis by multiplication by factors $10$,
$100$, and $1000$, respectively.  }
\label{fig:validation2}
%
\end{figure}

\subsection{Finite-difference method}
\label{sec:FDM}

For an independent validation, we also implemented a finite-difference
method for solving the original PDE problem for spherical shells.
The rotational invariance reduces the diffusion equation with $C_0 =
1$ to
\begin{align*}
& \partial_t C = D \bigl(\partial_r^2 + (2/r) \partial_r) C,  \qquad C(r,0) = 1,  \\
& \partial_r C|_{r = L} = 0,  \qquad (D\partial_r C - \kappa(t) C)|_{r = R} = 0.
\end{align*}
For a small length step $\sigma$, we set the timestep $\delta =
\sigma^2/(2D)$ and discretize this problem as
\begin{equation}
C(r,t+\delta) = C(r,t) + \frac12 \mathbf{L} C(r,t),
\end{equation}
where $\mathbf{L}$ is the dimensionless discrete diffusion operator,
with the elements
\begin{align*}
& \mathbf{L}_{j,j} = -2,  \quad \mathbf{L}_{j,j\pm 1} = 1 \pm \sigma/r_j   \quad (1 < j < N),\\
& \mathbf{L}_{1,1} = -1 - p_t - (1-p_t)\sigma /r_1, \quad  \mathbf{L}_{1,2} = 1 + \sigma/r_1, \\
& \mathbf{L}_{N,N} = -1 + \sigma/r_N, \quad  \mathbf{L}_{N,N-1} = 1 - \sigma/r_N,
\end{align*}
where $r_j = R + \sigma j$ with $j = 1,2,\ldots,N$ are the interior
points of the interval $(R,L)$, $N+1 = (L - R)/\sigma$, and $p_t = 1 -
1/(1 + \sigma \kappa(t)/D)$ (i.e., $p_t$ is given by
Eq. (\ref{eq:pt_exact}) with $\sigma$ instead of $\ve$).  This
discrete Laplacian incorporates the Neumann boundary condition at $r
=L$, which reads as $C(L,t) = C(L-\sigma,t)$, and the Robin boundary
condition at $r = R$, which reads as $C(R+\sigma,t) = C(R,t)(1 +
\sigma \kappa(t)/D)$.  Note that $\mathbf{L}$ depends on time through
$p_t$.  For validation analysis presented below, we used $\sigma =
0.02$.  Finally, the volume-averaged survival probability is obtained
by integrating $C(r,t)$ over $r$:
\begin{equation}
S(t) = \frac{3}{L^3-R^3} \int\limits_R^L dr \, r^2 \, C(r,t),
\end{equation}
which is evaluated with trapezoidal weights.   In turn, the PDF
was obtained by combining Eqs. (\ref{eq:Robin}, \ref{eq:J_def},
\ref{eq:Ht_Jt}) as
\begin{equation}
H(t) = \frac{3R^2}{L^3-R^3} \,\kappa(t) \, C(R,t) .
\end{equation}


\subsection{Numerical validation}
\label{sec:validation}

In this section, we apply the above numerical methods to compute the
PDFs $H(t)$ and $\He(t)$ for the power-law model
(\ref{eq:kappa_model}).  This comparison is used for cross-validation
of these methods.  We stress that we do not aim at performing a
systematic error analysis of these methods, nor comparing their
computational efficiency.

Figure \ref{fig:validation} shows the PDF $H(t)$ for time-dependent
reactivity $\kappa(t)$ obtained by two Monte Carlo implementations and
an independent finite-difference method: (i) direct Monte Carlo
simulations (Appendix \ref{sec:direct}), shown by circles; (ii) Monte
Carlo simulations of the functional $L_t$ (Appendix
\ref{sec:MC_proba}), shown by crosses; and (iii) results of
finite-difference method (Appendix \ref{sec:FDM}), shown by lines.  We
recall that all three methods solve the same problem so that their
predictions should be identical, within numerical errors.  This is
confirmed by Fig. \ref{fig:validation}, which shows excellent
agreement between three methods for all considered parameters.

Figure \ref{fig:validation2} compares the encounter-based Monte Carlo
simulations (Appendix \ref{sec:MC_encounter}) with the numerical
evaluation of the exact integral representation (\ref{eq:Ht_Ut})
(Appendix \ref{sec:shell}).  The two approaches show excellent
agreement over the considered time range.  We also compare these
results with the long-time predictions (\ref{eq:He_power}).  For
$\alpha=0$ and $\alpha=1$, the predicted power-law behavior is
accurately reproduced.  For $\alpha = -0.5$, the stretched-exponential
approximation captures the leading decay mechanism, but subleading
corrections may remain significant over the accessible time range, as
discussed in Appendix \ref{sec:Along}.  For $\alpha = -2$, the leading
exponential decay rate is controlled by the principal Dirichlet
eigenvalue, whereas the prefactor and subleading corrections are not
determined here; for this reason, we do not show this behavior.

\end{document}